\documentstyle[11pt]{article}

\newcommand{\sect}[1]{\setcounter{equation}{0}\section{#1}}

\newcommand{\EQ}{\begin{equation}}
\newcommand{\EN}{\end{equation}}
\newcommand{\bea}{\begin{eqnarray}}
\newcommand{\ena}{\end{eqnarray}}

\renewcommand{\b}{\beta}
\renewcommand{\d}{\delta}

\newcommand{\pa}{\partial}

\newcommand{\G}{\Gamma}

\newcommand{\e}{\epsilon}

\renewcommand{\l}{\lambda}

\newcommand{\m}{\mu}

\newcommand{\r}{\rho}

\begin{document}

\topmargin 0pt
\oddsidemargin 5mm

\renewcommand{\Im}{{\rm Im}\,}
\newcommand{\NP}[1]{Nucl.\ Phys.\ {\bf #1}}
\newcommand{\PL}[1]{Phys.\ Lett.\ {\bf #1}}
\newcommand{\NC}[1]{Nuovo Cimento {\bf #1}}
\newcommand{\CMP}[1]{Comm.\ Math.\ Phys.\ {\bf #1}}
\newcommand{\PR}[1]{Phys.\ Rev.\ {\bf #1}}
\newcommand{\PRL}[1]{Phys.\ Rev.\ Lett.\ {\bf #1}}
\newcommand{\MPL}[1]{Mod.\ Phys.\ Lett.\ {\bf #1}}
\renewcommand{\thefootnote}{\fnsymbol{footnote}}
\newpage
\begin{titlepage}
\vspace{2cm}
\begin{center}
{\bf{{\Large TOPOLOGICAL CORRELATION FUNCTIONS IN}}} \\
{\bf{{\Large MINKOWSKI SPACETIME}}} \\
\vspace{2cm}
{\large S. Penati and D. Zanon} \\
{\em Dipartimento di Fisica dell' Universit\`{a} di Milano and} \\
{\em INFN, Sezione di Milano, I-20133 Milano, Italy}\\
\end{center}
\vspace{2cm}
\centerline{{\bf{Abstract}}}
\vspace{.5cm}

We consider a class of non-unitary Toda theories based on the Lie
superalgebras $A^{(1)}(n,n)$ in
two-dimensional Minkowski spacetime, which can be twisted into a
topological sector. In particular
we study the simplest example with $n=1$ and real fields, and
show how this theory can be treated as
an integrable perturbation of the $A(1,0)$ superconformal model.
We construct the conserved currents and
the vertex operators which are chiral primary fields in the
conformal theory. We then define the chiral ring of the
$A^{(1)}(1,1)$ Toda theory and compute topological correlation
functions in the twisted sector. The calculation is performed
using a $N=2$ off--shell superspace approach.

\vfill
\noindent
IFUM-476-FT\hfill {July 1994}
\end{titlepage}
\renewcommand{\thefootnote}{\arabic{footnote}}
\setcounter{footnote}{0}
\newpage

\sect{Introduction}

Toda field theories in two spacetime dimensions are quantum
integrable systems
whose affine version realizes an off--critical
deformation of the corresponding conformal non--affine model. The
on--shell properties
of these theories
are well understood by now: the exact S--matrices have been
constructed \cite{b1}, being elastic
and two--body factorizable as a consequence of the existence of
higher--spin conserved currents \cite{b2}.

Off--shell the situation is not so clear: form factors and correlation
functions
have been computed only for few specific models \cite{b3} and in this case
integrability
doesn't seem to play the same relevant role as in the on--shell
counterpart.

The addition of supersymmetry might provide better hopes for the
construction of
completely solvable models. In particular it is well known that
chiral and
antichiral Green's functions of any $N=2$ supersymmetric theory
are spacetime
independent, a drastic simplification if one were to attempt their
complete
determination. These classes of Green's functions are the ones
which appear as
physical correlators in the twisted topological version \cite{b4}
of the $N=2$ supersymmetric
theory and for some specific choices of the superpotential they
have been computed exactly \cite{b5,b6}.

In this paper we focus on the calculation of chiral correlation
functions for the
class of $N=2$ supersymmetric Toda theories in Minkowski
spacetime which are based
on the $A^{(1)}(n,n)$ superalgebras \cite{Ol,b7,b8}.
These are nonunitary systems that can be
twisted into a topological, BRST invariant sector \cite{b9} where ghost--like
fields are
eliminated from the physical spectrum. There survives instead an
infinite number
of solitonic configurations \cite{b9} which at the quantum level can be
realized as
BRST invariant vertex operators. The identification of the chiral ring
and the study
of topological correlators are the main issues that we address.

In section 2 we briefly present the $N=2$ supersymmetric
formulation of the
$A^{(1)}(n,n)$ Toda field theory action. Then we concentrate on the
simplest
example with $n=1$ and real fields. The conserved currents and the
corresponding
charges are computed in Section 3. There we show how the
$A^{(1)}(1,1)$ affine
Toda theory can be obtained adding relevant
perturbation terms to the
superconformal $A(1,0)$ model. Section 4 is devoted to the
definition and
construction of the chiral ring: it consists in an infinite set of chiral
primary
vertex operators which generate physical states acting on the
vacuum. The
spectrum is infinite since we are dealing with a nonunitary theory.
Local
and nonlocal primary fields are present corresponding to classical
solutions
with trivial boundary conditions or solitonic ones respectively. They
are
constructed perturbatively with respect to the superconformal
$A(1,0)$
model and classified by their scale dimensions, the $O(1,1)$ charge
and the
solitonic charge. The topological
version of the theory is reviewed in Section 5, while the
computation of
topological correlation functions is the subject of Section 6.
We analyze how
the requirement of $O(1,1)$ charge conservation can be
implemented
consistently while maintaining background charge balance. We also
show that
spacetime indendence of the topological correlators reduces the
computation of
all of them to the calculation of one--point correlation functions. In
Section 7 we perform the explicit calculation for the $c=1$ model
to zero order in the relevant perturbation. We use superspace techniques which
not only
simplify the algebra
but are actually necessary for the correct introduction of screening
operators which are marginal perturbations of the free field theory.
Finally in Section 8 we present our conclusions.
Notations and conventions are listed
in Appendix A, while some details of the computation of the
correlation
function are collected in Appendix B.

\sect{The $A^{(1)} (n,n)$ affine Toda field theory}

Among the class of $N=1$ supersymmetric Toda field theories the
ones
associated to the $A^{(1)}(n,n)$ superalgebras admit a second
supersymmetry \cite{b10}
and they can be formulated in terms of $2n$, $N=2$ complex
superfields
$\Psi_i^{(\pm)}$, $\bar{\Psi}_i^{\pm}$ whose components are
$\Psi_i^{+} \to (\phi^{+}_i, \psi^{+}_i, \bar{\psi}^{+}_i, F^{+}_i
)$,
$\bar{\Psi}_i^{-} \to (\phi^{-}_i, \psi^{-}_i, \bar{\psi}^{-}_i,
F^{-}_i )$ and $\bar{\Psi}_i^{+} = \Psi_i^{+ \ast}$,
$\Psi_i^{-} = \bar{\Psi}_i^{- \ast}$
(conventions on $N=2$ superspace are listed in Appendix A).
The superspace action in two-dimensional Minkowski spacetime is
given by
\EQ
{\cal S} = \frac{1}{2\pi}~\left\{ \int d^2z d^4\theta ~K_{ij} \left[ \Psi_i^{+}
\bar{\Psi}_j^{-} + \Psi_i^{-} \bar{\Psi}_j^{+} \right] +
\frac{1}{ \beta^2} \int d^2z
d^2\theta~ W(\Psi)  + \frac{1}{ \beta^2}
\int d^2z d^2\bar{\theta}~ W(\bar{\Psi})\right\}
\label{1}
\EN
where the superpotential is
\EQ
W(\Psi) = \sum_{i=1}^n e^{\beta \Psi_i^{+}} + g_+ e^{-\b \sum_{i=1}^{n}
\Psi_i^{+}} + \sum_{i=1}^n e^{\beta \Psi_i^{-}} + g_- e^{-\b
\sum_{i=1}^{n}
\Psi_i^{-}}
\EN
$\beta$ is a coupling constant and $K_{ij}$ is the $n \times n$
matrix
\EQ
K=
\left(\begin{array}{cccccc}
1 & 1 & 1 & 1 & \cdots & ~ \\
1 & 0 & 0 & 0 & \cdots & ~ \\
1 & 0 & 1 & 1 & \cdots & ~ \\
1 & 0 & 1 & 0 & \cdots & ~ \\
\cdots & \cdots & \cdots & \cdots & \cdots & ~ \\
\cdots & \cdots & \cdots & \cdots & \cdots & ~
\end{array}\right)
\quad
K^{-1}=
\left(\begin{array}{cccccc}
0 & 1 & 0 & \cdots & ~ & ~ \\
1 & 0 & -1 & 0 & \cdots & ~ \\
0 & -1 & 0 & 1 & 0 & \cdots \\
{}~ & ~ & ~ & \cdots & ~ & ~ \\
{}~ & ~ & ~ & ~ & 0 & (-1)^n \\
{}~ & ~ & ~ & ~ & (-1)^n & (-1)^{n+1}
\end{array}\right)
{}~~~~~~~~~~~
\EN
We have included two coupling constants $g_+$ and $g_-$ in the
superpotential
for later convenience. The $A^{(1)}(n,n)$ theory is obtained by
setting
$g_+=g_-=1$, whereas for $g_-=1, ~g_+=0$ the superpotential
reduces to the
one for the $N=1$ superconformal invariant $A(n,n)$ theory and for
$g_+=g_-=0$ one obtains the $N=2$ superconformal $A(n,n-1)$
Toda. At the
component level, after elimination of the auxiliary fields, the action
can be
reexpressed as
\bea
{\cal S} &=& \frac{1}{2\pi}
\int d^2z \left\{ K_{ij} \left[ \pa \phi_i^{-} \bar{\pa}\phi_j^{+}
+ \frac{i}{2} \psi_i^{-} \bar{\pa} \psi_j^{+} -\frac{i}{2}
\bar{\psi}_i^{-} \pa \bar{\psi}_j^{+} \right]
- \frac{\pa V}{\pa \phi_i^{+}} K_{ij}^{-1}
\frac{\pa { V}}{\pa \phi_j^{-}} \right. \nonumber \\
&~~&~~~~~~\qquad \left. + \frac12 \bar{\psi}_i^{+} \psi_j^{+}
\frac{\pa^2 V}
{\pa \phi_i^{+} \pa \phi_j^{+}}  +
\frac12 \bar{\psi}_i^{-} \psi_j^{-} \frac{\pa^2 { V}}
{\pa \phi_i^{-} \pa \phi_j^{-}} ~+~{\rm h.c.} \right\}
\label{129}
\ena
where
\EQ
V(\phi^{\pm} )= \frac{1}{\b^2} \left[ \sum_{i=1}^n e^{\b\phi_i^{\pm}} +
g_{\pm} e^{-\b \sum_{i=1}^n \phi_i^{\pm}} \right]
\label{124}
\EN
These theories are integrable and in Refs. \cite{b7,b8} the first non trivial
higher--spin quantum currents were constructed to all orders in
perturbation theory. Moreover they are not unitary since the
matrix $K_{ij}$ is not positive definite.

The action in eq. (\ref{1}) is invariant under a complexified $N=2$
algebra
generated by the supersymmetry operators
\bea
Q^+ &=& \frac{1}{\sqrt{2}} \left( \frac{\partial}{\partial \theta_+}
- i
\bar{\theta}_+
\partial \right) \qquad
\bar{Q}^+ = \frac{1}{\sqrt{2}} \left( \frac{\partial}{\partial \theta_-} + i
\bar{\theta}_- \bar{\partial} \right)
\nonumber \\
Q^- &=& \frac{1}{\sqrt{2}} \left(
\frac{\partial}{\partial \bar{\theta}_+} - i \theta_+ \partial \right)
\qquad \bar{Q}^- = \frac{1}{\sqrt{2}} \left( \frac{\partial}{\partial
\bar{\theta}_-} + i\theta_- \bar{\partial} \right)
\ena
which satisfy $\{ Q^{+},Q^{-} \} = -i\pa$, $\{ \bar{Q}^{+},\bar{Q}^{-} \} =
i\bar{\pa}$.
In these models the usual $U(1)$
invariance of $N=2$ supersymmetry is extended to a $U(1) \otimes
O(1,1)$ invariance. As shown in Ref. \cite{b9} this allows to perfom a
topological twist of the theory directly in Minkowski space by
combining
the $O(1,1)$ Lorentz group with this extra $O(1,1)$ invariance.

When the coupling constant $\beta$ is imaginary the classical
equations of
motion from the action (2.4) with $g_+=g_-=1$
admit solitonic solutions which have been
studied in Ref. \cite{b9}. In particular the soliton spectrum survives
the twist
and the topological sector has the same solitonic
content as
the bosonic $a_n^{(1)}$ theory \cite{b11}.

\sect{The model}

We restrict now our attention to the simplest model in the above
mentioned class, the
$A^{(1)}(1,1)$ theory with coupling constant $i\beta$ in the
exponentials.
Furthermore we choose the scalar fields $\phi^{\pm}$ to be real
and $\psi^{\pm}$, $\bar{\psi}^{\pm}$ to be the chiral and
antichiral
components of two Majorana fermions ($\psi^{\pm\ast}=
\psi^{\pm}$,
$\bar{\psi}^{\pm\ast} = -\bar{\psi}^{\pm}$).
As discussed in Ref. \cite{b9} these reality conditions
do not affect the $O(1,1)$ invariance which allows to perform the
twist into the topological sector. However the $U(1)$ invariance is
lost and
the theory does not possess a genuine $N=2$ supersymmetry.
The action becomes
\bea
{\cal S} &=& \frac{1}{2\pi} \int d^2z \left[ \partial \phi^{+}
\bar{\partial}
\phi^{-} + \frac{i}{2} \psi^{+} \bar{\partial} \psi^{-} -
\frac{i}{2} \bar{\psi}^{+} \partial \bar{\psi}^{-} \right.
\nonumber \\
&+& \left. \frac{1}{\beta^2} e^{i\b (\phi^{+} +\phi^{-})}
-\frac{1}{\b^2} g_- e^{i\b (\phi^{+} -\phi^{-})} - \frac{1}{\b^2}
g_+
e^{i\b(\phi^{-} -\phi^{+})} +\frac{1}{\b^2} g_+g_-
e^{-i\b(\phi^{+}+\phi^{-})} \right. \nonumber\\
&-& \left. \frac12 \bar{\psi}^{+} \psi^{+} \left( e^{i\b\phi^{+}} +
g_+
e^{-i\b\phi^{+}}\right) -\frac12 \bar{\psi}^{-} \psi^{-} \left(
e^{i\b\phi^{-}} +g_-e^{-i\b\phi^{-}} \right) \right] \nonumber
\\
& \equiv & {\cal S}_0 + {\cal S}_{\rm pert}
\label{2}
\ena
where
\EQ
{\cal S}_0 \equiv \frac{1}{2\pi} \int d^2z~ \left[ \partial \phi^{+}
\bar{\partial} \phi^{-} + \frac{i}{2} \psi^{+} \bar{\partial} \psi^{-} -
\frac{i}{2} \bar{\psi}^{+} \partial \bar{\psi}^{-} + V_0 \right]
\label{act}
\EN
with
\EQ
V_0 \equiv \frac{1}{\beta^2} e^{i\b (\phi^{+} +\phi^{-})}
-\frac12 \bar{\psi}^{+} \psi^{+} e^{i\b\phi^{+}}
-\frac12 \bar{\psi}^{-} \psi^{-} e^{i\b\phi^{-}}
\label{pot}
\EN
is the action of the $A(1,0)$ theory.
The bosonic and fermionic fields satisfy equal--time commutation
relations
\bea
[\phi^{\pm}(x^0,x^1), \dot{\phi}^{\mp}(y^0,y^1)]_{\mid x^0=y^0}
&=& 4\pi
i ~\delta (x^1-y^1) \nonumber \\
\{ \psi^{\pm}(x^0,x^1),\psi^{\mp}(y^0,y^1)\}_{\mid x^0=y^0} &=&
4\pi
\sqrt{2} ~\delta(x^1-y^1) \nonumber \\
\{ \bar{\psi}^{\pm}(x^0,x^1),\bar{\psi}^{\mp}(y^0,y^1)\}_{\mid
x^0=y^0}
&=& -4\pi \sqrt{2} ~\delta(x^1-y^1)
\label{3}
\ena
In the following we will drop the subscript $x^0 =y^0$, all
commutators
being evaluated at equal times.

Despite the lack of $U(1)$ invariance the theory can be formulated
in
$N=2$ superspace. In terms of two superfields $\Psi^{+}$ and
$\Psi^{-}$,
chiral and antichiral respectively, with components
$\Psi^{\pm} \to (\phi^{\pm}, \psi^{\pm}, \bar{\psi}^{\pm},
F^{\pm} )$ the action in eq. (\ref{2}) can be written as
\bea
{\cal S} &=& \frac{1}{2\pi} \left\{ \int d^2z d^4\theta~ \Psi^{+} \Psi^{-}
+\frac{1}{\b^2} \int d^2zd^2\theta \left[ e^{i\b \Psi^{+}} + g_+
e^{-i\b
\Psi^{+}} \right] \right. \nonumber \\
&~&~~~~~~~+ \left. \frac{1}{\b^2} \int d^2zd^2\bar{\theta}
\left[ e^{i\b \Psi^{-}} + g_- e^{-i\b \Psi^{-}} \right] \right\}
\label{311}
\ena
Later we will use the superspace formulation as a suitable device to
compute
correlation functions.

The symmetries of the model are generated by the following
conserved
currents: the spin--2 stress--energy tensor, the spin--1 $O(1,1)$
generator
and the two spin--$\frac{3}{2}$ supersymmetries.
According to the definitions given in Appendix A, we have
\bea
T &=& -\partial \phi^{+} \partial \phi^{-} - \frac{i}{4} \psi^{-}
\partial \psi^{+} - \frac{i}{4} \psi^{+} \partial \psi^{-}
-\frac{i}{2\b} \partial^2\phi^{+}  -\frac{i}{2\b} \partial^2\phi^{-}
\nonumber \\
J &=& \frac{i}{2} \psi^{+} \psi^{-} +\frac{i}{\b}\partial \phi^{+}
-\frac{i}{\b} \partial \phi^{-} \nonumber \\
G^{+} &=& -\psi^{+} \partial \phi^{-} -\frac{i}{\b} \partial
\psi^{+}
\nonumber \\
G^{-} &=& i\psi^{-} \partial \phi^{+} -\frac{1}{\b} \partial
\psi^{-}
\label{112}
\ena
which satisfy the following conservation equations \cite{b8}
\bea
\bar{\pa} J &=& \pa \left[ \frac{i}{2} \bar{\psi}^{-} \bar{\psi}^{+}
+\frac{i}{\b}\bar{\pa} \phi^{+} -\frac{i}{\b}\bar{\pa} \phi^{-}
\right]
\equiv -\pa \tilde{J}
\nonumber \\
\bar{\pa} G^{+} &=& \pa \left( \frac{2}{\b}g_- \bar{\psi}^{-}
e^{-i\b\phi^{-}} \right) -
\frac{1}{\b} \pa
\left[ i\bar{\pa} \psi^{+} + \bar{\psi}^{-}
\left( e^{i\b\phi^{-}} +g_-e^{-i\b\phi^{-}}
\right) \right] \equiv -\pa \tilde{G}^{+} \nonumber \\
\bar{\pa} G^{-} &=& \pa \left( -\frac{2}{\b}ig_+ \bar{\psi}^{+}
e^{-i\b\phi^{+}} \right) + \frac{i}{\b} \pa
\left[ i\bar{\pa} \psi^{-} + \bar{\psi}^{+} \left( e^{i\b\phi^{+}} +
g_+e^{-i\b\phi^{+}} \right) \right] \equiv -\pa \tilde{G}^{-}
\nonumber \\
\bar{\pa} T &
=& \pa \left[ \frac{1}{\b^2} \left( g_-
e^{i\b (\phi^{+} -\phi^{-})}+g_+ e^{i\b(\phi^{-} -\phi^{+})}
-2g_+g_-  e^{-i\b(\phi^{+}+\phi^{-})} \right) \right. \nonumber
\\
{}~~~~~&~~& ~~~~~\left. +\frac12 g_- \bar{\psi}^{-} \psi^{-}
e^{-i\b\phi^{-}}
+ \frac12 g_+  \bar{\psi}^{+} \psi^{+} e^{-i\b\phi^{+}} \right]
\nonumber \\
{}~~~~~&~~& -\frac{i}{4} \pa \left[ \psi^{+} \bar{\pa} \psi^{-} -
i \psi^{+} \bar{\psi}^{+}
\left( e^{i\b\phi^{+}} + g_+ e^{-i\b\phi^{+}} \right) \right]
\nonumber \\
{}~~~~~&~~& -\frac{i}{4} \pa \left[ \psi^{-} \bar{\pa} \psi^{+} -
i \psi^{-} \bar{\psi}^{-}
\left( e^{i\b\phi^{-}} + g_-e^{-i\b\phi^{-}} \right) \right]
\nonumber \\
{}~~~~~&~~& -\frac{i}{2\b} \pa \left[ \pa \bar{\pa} \phi^{-}
-\frac{i}{\b}
\left( e^{i\b\phi^{+}} +
g_+ e^{-i\b\phi^{+}} \right) \left( e^{i\b\phi^{-}} - g_-e^{-
i\b\phi^{-}}
\right) \right. \nonumber \\
{}~~~~~&~~& ~~~~~\left. + \frac{i\b}{2}
\bar{\psi}^{+} \psi^{+} \left( e^{i\b\phi^{+}} -g_+ e^{-
i\b\phi^{+}}
\right) \right] \nonumber \\
{}~~~~~&~~& -\frac{i}{2\b} \pa \left[ \pa \bar{\pa} \phi^{+}
-\frac{i}{\b}
\left( e^{i\b\phi^{+}} -
g_+ e^{-i\b\phi^{+}} \right) \left( e^{i\b\phi^{-}} + g_-e^{-
i\b\phi^{-}}
\right) \right. \nonumber \\
{}~~~~~&~~& ~~~~~\left. + \frac{i\b}{2}
\bar{\psi}^{-} \psi^{-} \left( e^{i\b\phi^{-}} -g_- e^{-i\b\phi^{-}}
\right) \right]
\equiv -\pa \tilde{T}
\label{4}
\ena
We also have
\bea
\bar{T} &=& -\bar{\partial} \phi^{+} \bar{\partial} \phi^{-} +
\frac{i}{4} \bar{\psi}^{-} \bar{\partial} \bar{\psi}^{+} +
\frac{i}{4} \bar{\psi}^{+} \bar{\partial} \bar{\psi}^{-}
-\frac{i}{2\b} \bar{\partial}^2\phi^{+}  -\frac{i}{2\b}
\bar{\partial}^2
\phi^{-} \nonumber \\
\bar{J} &=& \frac{i}{2} \bar{\psi}^{+} \bar{\psi}^{-} -
\frac{i}{\b}\bar{\partial} \phi^{+} +\frac{i}{\b} \bar{\partial}
\phi^{-} \nonumber \\
\bar{G}^{+} &=& -\bar{\psi}^{+} \bar{\partial} \phi^{-} -
\frac{i}{\b} \bar{\partial} \bar{\psi}^{+}
\nonumber \\
\bar{G}^{-} &=& i\bar{\psi}^{-} \bar{\partial} \phi^{+} -
\frac{1}{\b} \bar{\partial} \bar{\psi}^{-}
\label{113}
\ena
for which similar conservation equations hold. (We note that
$\tilde{J}=\bar{J}$ and ${\tilde{\bar{J}~}}=J$).

Setting $g_+=g_-=0$ in eq. (\ref{4}), the right--hand sides
vanish on--shell since the $A(1,0)$ theory is $N=2$
superconformal
invariant and on--shell conservations hold separately in the
holomorphic
and antiholomorphic sectors.
In any case terms proportional to the equations of motion
are necessary whenever the off--shell invariance of the action
needs be
exhibited.

We note that in eqs. (\ref{112}), (\ref{113}) improvement terms,
i.e. total
derivative terms, have been included in order to implement the
correct
holomorphic currents in the $A(1,0)$ theory. In particular the term
$-\frac{i}{2\b}(\pa^2 \phi^{+} + \pa^2\phi^{-})$ in $T$ and the
corresponding one in $\bar{T}$ signal the presence
of a background charge ($\frac{1}{\b}$, $\frac{1}{\b}$)
coupled to the ($\phi^{+}$, $\phi^{-}$) fields at infinity. Moreover
it is easy to check that the central charge for the improved stress--energy
tensor is
$c =3-\frac{6}{\b^2}$.

In general for a spin--$s$ current $J^{(s)}$ which satisfies the
conservation equation $\bar{\pa} J^{(s)} + \pa \tilde{J}^{(s)}=0$, the
corresponding charge is $\int \frac{dx}{2\pi i \sqrt{2}}
 (J^{(s)} + \tilde{J}^{(s)})$.
Therefore from eq. (\ref{4}) and its counterpart in the
antiholomorphic sector
we can construct the conserved charges of the model. In particular,
the supersymmetry charges are
\bea
G^{+}_{-\frac12} &=& \int
\frac{dx}{2\pi i \sqrt{2}} \left[ -\psi^{+}
\pa \phi^{-} + \frac{1}{\b} \bar{\psi}^{-} \left( e^{i\b\phi^{-}} -
g_- e^{-i\b\phi^{-}} \right)  \right]   \nonumber \\
G^{-}_{-\frac12} &=& \int
\frac{dx}{2\pi i \sqrt{2}} \left[ i\psi^{-}
\pa \phi^{+} - \frac{i}{\b} \bar{\psi}^{+} \left( e^{i\b\phi^{+}} -
g_+ e^{-i\b\phi^{+}} \right) \right]   \nonumber \\
\bar{G}^{+}_{-\frac12} &=& \int
\frac{dx}{2\pi i \sqrt{2}} \left[ -\bar{\psi}^{+}
\bar{\pa} \phi^{-} + \frac{1}{\b} \psi^{-} \left( e^{i\b\phi^{-}} -
g_- e^{-i\b\phi^{-}} \right)  \right]   \nonumber \\
\bar{G}^{-}_{-\frac12} &=& \int
\frac{dx}{2\pi i \sqrt{2}} \left[ i\bar{\psi}^{-}
\bar{\pa} \phi^{+} - \frac{i}{\b} \psi^{+} \left( e^{i\b\phi^{+}} -
g_+ e^{-i\b\phi^{+}} \right) \right]
\label{111}
\ena
In terms of the charges the conservation equations
satisfied by the supersymmetry currents can be written as follows
\bea
\bar{\pa} G^{+} &=& -\frac{2}{\b^2} ~g_- \pa \left[
\bar{G}^{-}_{-\frac12}, e^{-i\b \phi^{-}} \right]  \nonumber \\
\bar{\pa} G^{-} &=& \frac{2}{\b^2} ~g_+ \pa \left[
\bar{G}^{+}_{-\frac12}, e^{-i\b \phi^{+}} \right]  \nonumber \\
\pa \bar{G}^{+} &=& -\frac{2}{\b^2} ~g_- \bar{\pa} \left[
G^{-}_{-\frac12}, e^{-i\b \phi^{-}} \right]    \nonumber \\
\pa \bar{G}^{-} &=&  \frac{2}{\b^2}~ g_+ \bar{\pa} \left[
G^{+}_{-\frac12}, e^{-i\b \phi^{+}} \right]
\ena
These relations are exact to all orders in perturbation theory. For
$g_+=0$ they agree with the result given in Ref. \cite{b12}
where the perturbation corresponding to antichiral fields
was neglected.

The term ${\cal S}_{\rm pert}$ in eq. (3.1)
which must be added to the $A(1,0)$
superconformal action $S_0$ in order to obtain the affine
$A^{(1)}(1,1)$
theory, can be reexpressed in terms of the supersymmetry charges
defined in
eq. (\ref{111}).
More precisely, starting from ${\cal S}_0$ in eq. (\ref{act}), one
obtains the $A(1,1)$ Toda theory by adding
\EQ
{\cal S}_- = \frac{g_-}{4\pi\b^2} \int d^2z \left[ G^{-}_{-\frac12}~ ,
\left[
\bar{G}^{-}_{-\frac12}~, e^{-i\b\phi^{-}(z,\bar{z})} \right] \right]
\label{310}
\EN
where $G^{-}_{-\frac12}$, $\bar{G}^{-}_{-\frac12}$ are the
$A(1,0)$ supersymmetry charges.
As can be seen from eq. (\ref{111}) the presence of the
${\cal S}_-$
perturbation leaves $G^{-}_{-\frac12}$, $\bar{G}^{-}_{-\frac12}$
unchanged whereas a $g_-$--dependence arises in $G^{+}_{-
\frac12}$,
$\bar{G}^{+}_{-\frac12}$. Then the action for the affine
$A^{(1)}(1,1)$ Toda
is constructed by perturbing the $A(1,1)$ theory with
\EQ
{\cal S}_+ = -\frac{g_+}{4\pi\b^2} \int d^2z \left[ G^{+}_{-\frac12}~ ,
\left[ \bar{G}^{+}_{-\frac12}~, e^{-i\b\phi^{+}(z,\bar{z})} \right]
\right]
\label{10}
\EN
where in eq. (\ref{10}) $G^{+}_{-\frac12}$, $\bar{G}^{+}_{-
\frac12}$ are
the perturbed $A(1,1)$ charges. The ${\cal S}_+$ perturbation
modifies the
$G^{-}_{-\frac12}$, $\bar{G}^{-}_{-\frac12}$ charges as in eq.
(\ref{111}), leaving
$G^{+}_{-\frac12}$, $\bar{G}^{+}_{-\frac12}$ unchanged
in this second step. \\
{}~~~~~~~~~~~~~~~~\\

Classical solutions of the equations of motion from the
action (\ref{2}) are classified by the value of the topological
charges
\EQ
{\cal T}^{\pm} = \frac{\b}{2\pi} \int dx~
\frac{\pa
\phi^{\pm}}{\pa x} = \frac{\b}{2\pi}
\left[ \phi^{\pm}(+\infty)-\phi^{\pm}(-\infty)\right]
\label{5}
\EN
For $g_+, g_- \not= 0$ solitonic sectors are characterized by
different,
non zero values of ${\cal T}^{+}$ and ${\cal T}^{-}$.
 At the quantum level we promote the topological
charges ${\cal T}^{\pm}$ to be operators which, acting on physical
states,
give the solitonic number. States with different solitonic content
are
orthogonal.

We turn now to the definition of the chiral ring of
the $A^{(1)}(1,1)$ theory.

\sect{The chiral ring}

We study first the spectrum of primary fields for the affine
Toda theory. It consists in a set of local and nonlocal
vertex operators which are classified by three quantum
numbers: the scale dimensions, the $O(1,1)$ charge and the
topological
charge. We construct them perturbatively in $g_+$ and $g_-$
treating the
affine $A^{(1)}(1,1)$ theory as an integrable perturbation of the
$N=2$
superconformal $A(1,0)$ model. Therefore the quantum numbers of
the vertex
operators are computed with respect to the $A(1,0)$ charges
\bea
L_0 &=& \int
\frac{dx}{2\pi i \sqrt{2}} ~\frac{x}{\sqrt{2}}~ T \qquad \quad
\bar{L}_0 ~=~ \int
\frac{dx}{2\pi i \sqrt{2}} ~\frac{x}{\sqrt{2}} ~\bar{T} \nonumber \\
J_0 &=& \int \frac{dx}{2\pi i \sqrt{2}} ~J \qquad
\quad \qquad
\bar{J}_0 ~=~ \int \frac{dx}{2\pi i \sqrt{2}}
{}~\bar{J}
\label{100}
\ena
with $\bar{\pa}T=\bar{\pa}J=0$ and $\pa\bar{T}=\pa\bar{J}=0$
on--shell.

The local operators are given by
\EQ
\Phi_a^{\pm} = :e^{ia\b\phi^{\pm}}: \qquad \qquad {\rm any}~ a \in
{\cal R}
\label{8}
\EN
(In the following vertex operators are always normal ordered even
when
not explicitly indicated.)
Due to the lack of unitarity the spectrum is infinite and we do not
have
any restriction
on the value of $a$. Computing the commutators with the
holomorphic and antiholomorphic components of the stress--energy
tensor,
the $O(1,1)$ and the topological charges we obtain
\bea
\left[ L_0 , \Phi_a^{\pm} \right] &=&
\left[ \bar{L}_0 , \Phi_a^{\pm} \right] = \frac{a}{2} \Phi_a^{\pm}
\nonumber \\
\left[ J_0 , \Phi_a^{\pm} \right] &=&
- \left[ \bar{J}_0 , \Phi_a^{\pm} \right] = \mp a \Phi_a^{\pm}
\nonumber \\
\left[ {\cal T}^{\pm} ,\Phi_a^{\pm}\right] &=& 0
\ena
This shows that the local operators in eq. (\ref{8}) are characterized by
conformal weight $a$ and spin zero. Moreover the
holomorphic and antiholomorphic conformal dimensions are always
half of the
$O(1,1)$ charges for any $a$.

In order to construct the non--local sector of primary fields it
is convenient to introduce four non--local fields \cite{b13,b14}
\bea
\r^{\pm} &=& \frac12 \left[ \phi^{\pm} + \int_{-\infty}^{x} dy
{}~\dot{\phi}^{\pm} \right] \nonumber \\
\bar{\r}^{\pm} &=& \frac12 \left[ \phi^{\pm} - \int_{-\infty}^{x}
dy ~\dot{\phi}^{\pm} \right]
\ena
They satisfy
\bea
\bar{\pa} \r^{\pm} &=& \frac{1}{\sqrt{2}} \int_{-\infty}^{x} dy~
\frac{\pa V_0}{\pa \phi^{\mp}} \nonumber \\
\pa \bar{\r}^{\pm} &=&
-\frac{1}{\sqrt{2}} \int_{-\infty}^{x} dy~ \frac{\pa V_0}{\pa
\phi^{\mp}}
\ena
where $V_0$ is the $A(1,0)$ potential in eq. (\ref{pot}). From
eq. (\ref{3}) it is easy
to derive the equal--time commutators
for the creation and annihilation components
$\r^{\pm}_{\pm}$
\bea
\left[ \r^{\pm}_-(x^0,x^1),\r^{\mp}_+(y^0,y^1) \right]_{\mid
x^0=y^0} &=&
-\log[i \m(x^1-y^1-i\e)]  \nonumber \\
\left[ \bar{\r}^{\pm}_-(x^0,x^1),\bar{\r}^{\mp}_+(y^0,y^1)
\right]_{\mid
x^0=y^0} &=&
-\log[-i \m(x^1-y^1+i\e)] \nonumber \\
\left[ \r^{\pm}_-(x^0,x^1),\bar{\r}^{\mp}_+(y^0,y^1) \right]_{\mid
x^0=y^0}
&=& \left[ \r^{\pm}_+(x^0,x^1),\bar{\r}^{\mp}_-(y^0,y^1) \right]_{\mid
x^0=y^0} =-i\frac{\pi}{2}
\label{7}
\ena
where $\e$ and $\m$ are ultraviolet and infrared cutoffs respectively.
In terms of the fields in eq. (4.4) we can write
the most general non--local vertex operator as
\EQ
\Phi^{\pm}_{(a,b)} = e^{ia\b\r^{\pm} +ib\b\bar{\r}^{\pm}}
\EN
It reduces to the local vertex in eq. (\ref{8}) for $a=b$.
Using the equal--time commutation relations in eq. (\ref{7})
we can easily find the conditions that the non--local
operators need satisfy to be primary fields and then compute
their scale dimensions, the $O(1,1)$ and topological charges.
They have well--defined conformal
dimensions if
\EQ
\left[  V_0, \Phi^{\pm}_{(a,b)}
\right] = 0
\EN
i.e. they must
commute with
$e^{i\b\phi^{\mp}}$. Using results in Ref. \cite{b14} one shows
that this amounts to
\EQ
a-b = \frac{n}{\b^2} \qquad n \in {\cal Z} ~~~~~{\rm and} ~~~~
\begin{array} {cc}
\b^2 a > \frac{n-1}{2} & n~ {\rm odd} \\
\b^2 a > \frac{n-2}{2} & n~ {\rm even}
\end{array}
\label{98}
\EN
Finally we obtain the conformal dimensions, the $O(1,1)$ and the
topological charges
\bea
\left[ L_0 , \Phi^{\pm}_{(a,b)} \right] &=& \frac{a}{2}
\Phi^{\pm}_{(a,b)} \qquad ~~~~Ê\qquad
\left[ \bar{L}_0 , \Phi^{\pm}_{(a,b)} \right] = \frac{b}{2}
\Phi^{\pm}_{(a,b)} \nonumber \\
&&~~~~~~~~~~~~~~~~~~~~~~~~~~~~~~~~~~~~~~~ \nonumber \\
\label{tp}
\left[ J_0 , \Phi^{\pm}_{(a,b)} \right] &=& \mp a
\Phi^{\pm}_{(a,b)}
\qquad ~~\qquad
\left[ \bar{J}_0 , \Phi^{\pm}_{(a,b)} \right] = \pm b
\Phi^{\pm}_{(a,b)} \nonumber
\ena
\EQ
\left[ {\cal T}^{+},\Phi^{\pm}_{(a,b)} \right] = \left\{
\begin{array} {c}
0~~~~~~~~~~~~~~~ \\
\b^2 (a-b) \Phi^-_{(a,b)}
\end{array} \right. \qquad
\left[ {\cal T}^{-},\Phi^{\pm}_{(a,b)} \right] = \left\{
\begin{array} {c}
\b^2(a-b) \Phi^+_{(a,b)}\\
0~~~~~~~~~~~~~~~
\end{array} \right. \nonumber
\EN
with $\b^2 (a-b)=n$.
Thus the conformal weight of the operator is $\frac{a+b}{2}$
and the
spin $\frac{n}{2\b^2}$. For $n>0$ it describes solitons, whereas for
$n<0$
it creates antisolitonic states.

Given the spectrum of primary fields we can define now the chiral ring.
Since the theory we are dealing with is nonunitary, the definition
of the chiral ring is not unique (see Ref. \cite{b15}). We choose
to define it in the following way: left, right--chiral primaries satisfy
the condition
\EQ
G^{+}_{-\frac12} \Phi = \bar{G}^{+}_{-\frac12} \Phi = 0
\EN
where $\Phi$ is either a local or a non--local primary vertex operator.
Correspondingly left, right--antichiral fields are realized imposing
\EQ
G^{-}_{-\frac12} \Phi = \bar{G}^{-}_{-\frac12} \Phi = 0
\EN
We note that since the theory is interacting the rings are not
factorized into holomorphic and antiholomorphic sectors.
{}From the explicit expression of the supersymmetry charges it
follows that
chiral primary fields are of the form $\phi^{-}_{(a,b)}$
and antichiral fields are $\phi^{+}_{(a,b)}$ with $a$ and $b$
satisfying
(\ref{98}).

The physical states of the $A^{(1)}(1,1)$ theory in
the chiral and antichiral rings are constructed by acting
with the
vertex operators $\Phi_{(a,b)}$ on the $N=2$ supersymmetric
vacuum. The
states $|a,b\rangle = \Phi_{(a,b)} |0\rangle$ with $a \not= b$
correspond to classical solutions which exhibit a non--trivial
behavior at
infinity. In the $A(1,0)$ and $A(1,1)$ Toda theories where classical
solitonic solutions are not present, these states are eliminated by
the
extra conditions
\EQ
\left[ {\cal T}^{+} , \Phi_{(a,b)}^{-} \right] = 0 \qquad \qquad
\left[ {\cal T}^{-} , \Phi_{(a,b)}^{+} \right] = 0
\label{116}
\EN

\sect{Topological $A^{(1)}(1,1)$ theory}

The $O(1,1)$ symmetry of the affine Toda action in eq. (\ref{2}) can
be
combined with the Lorentz invariance to twist the theory into the
topological sector \cite{b9}. The spin content of the
topological theory is determined by the new Lorentz group $ L' = (L
\otimes O(1,1))_{\rm diag}$. Two equivalent twists are viable, the
only
difference being a spin interchange between
$\psi$ and $\bar{\psi}$. We choose the new spin
assignment
\bea
&&\psi^{+} ~~~~~s'=0 \qquad \qquad~~~ \bar{\psi}^{+} ~~~~~s'=0
\nonumber \\
&&\psi^{-} ~~~~~s'=1 \qquad \qquad ~~~ \bar{\psi}^{-} ~~~~~s'=-1
\ena
which corresponds to a  twist of the stress--energy tensor as $T' = T
+\frac12 \pa J$, $\bar{T}' = \bar{T} - \frac12 \bar{\pa} \bar{J}$
and $\tilde{T}' = \tilde{T} -\frac12 \bar{\pa}J$. It has central
charge equal
to zero.
With respect to the new spin assignment the supersymmetry
charges
$G^{+}_{-\frac12}$ and $\bar{G}^{+}_{-\frac12}$ have spin zero.
Therefore the nihilpotent combination
\bea
Q&=& iG^{+}_{-\frac12} +\bar{G}^{+}_{-\frac12} \\
&=&
\int dx
\frac{i}{2\pi\sqrt{2}}  \left[  \left( i\psi^{+}
\pa \phi^{-} + \bar{\psi}^{+} \bar{\pa} \phi^{-}\right) -
\frac{1}{\b} \left( \psi^{-}
+ i\bar{\psi}^{-}\right)  \left( e^{i\b\phi^{-}} - g_-
e^{-i\b\phi^{-}} \right) \right] \nonumber
\label{412}
\ena
is the BRST charge of the topological theory.
Moreover it is easy to check that
the twisted stress--energy tensor is $Q$--exact
\bea
T' &=& \left\{ Q, \frac12 \psi^{-} \pa \phi^{+} +\frac{i}{2\b}
\pa \psi^{-}
\right\} \nonumber \\
\bar{T}' &=& \left\{ Q, -\frac{i}{2} \bar{\psi}^{-} \bar{\pa}
\phi^{+}
+ \frac{1}{2\b} \bar{\pa} \bar{\psi}^{-}\right\} \nonumber \\
\tilde{T}' &=& \left\{ Q, -\frac{i}{\b} g_+ \psi^{+} e^{-i\b\phi^{+}}
\right\}
\ena
The BRST transformations on the fields are
\bea
\delta \phi^{+} &=& \eta[ i \psi^{+} +\bar{\psi}^{+}] \nonumber
\\
\delta \phi^{-} &=& 0  \nonumber \\
\delta \psi^{+} &=&  -2\eta
\frac{\pa V}{\pa \phi^{-}} \nonumber \\
\delta \psi^{-} &=&- 2\eta \pa \phi^{-} \nonumber \\
\delta \bar{\psi}^{+} &=& 2\eta i \frac{\pa V}{\pa
\phi^{-}} \nonumber \\
\delta \bar{\psi}^{-} &=&  -2i\eta \bar{\pa} \phi^{-}
\ena
where $\eta$ is a real spin--0 parameter and $V(\phi^{-})$ is the
potential in eq. (\ref{124}) with coupling constant $i\b$ in the
exponentials.
The physical spectrum of the topological theory is defined by the
condition
$Q |{\rm phys} \rangle =0$. This implies that only BRST--invariant
vertex operators generate physical states when acting on the
topological
vacuum ($Q |0\rangle = 0$). From the explicit expression of $Q$
in terms of the supersymmetry charges it follows that
the physical spectrum of the
topological theory coincides with the chiral ring of the untwisted
theory.
The unperturbed physical states are then generated by local
primaries
$e^{ia\b\phi^{-}}$ and $n$--soliton operators
$e^{ia\b\r^{-}+ib\b\bar{\r}^{-}}$, where $a$, $b$ satisfy the
conditions in eq. (\ref{98}).

We note that $\frac{\pa V}{\pa \phi^{-}} $ is
$Q$--exact, being proportional to the fermionic $Q$--transformations.
Thus $\phi^{-}$
is cohomological trivial except at the critical points of the potential.
We define our theory perturbatively around one of these points.

So far we have constructed the unperturbed physical spectrum of
the
topological theory. The perturbation modifies the chiral ring
introducing a non--trivial dependence on the couplings.
The perturbed chiral ring can be obtained by computing
perturbatively the structure constants (three points correlation
functions)
of the ring. The computation of correlation functions will be the
subject
of the next section.

\sect{Topological correlation functions}

We consider the most general
correlation function in the topological sector  of the $A^{(1)}(1,1)$
Toda theory
\EQ
{\cal F}_N \equiv \langle 0 | \Phi^{-}_{(a_1,b_1)}(x_1,t_1)
\Phi^{-}_{(a_2,b_2)}(x_2,t_2) \cdots
\Phi^{-}_{(a_N,b_N)}(x_N,t_N) |0\rangle_{\rm top}
\label{9}
\EN
for a string of $N$ local or solitonic vertex operators where
$|0\rangle$
is the topological vacuum. Exploiting the BRST invariance of the
chiral fields at fixed $t$
and using the relation $[Q,H]=0$ where $H$ is the hamiltonian of the
theory, it is immediate to check that the expression (\ref{9}) is
invariant under BRST transformations.
Consequently the topological correlation functions
are numbers, independent of positions and times.
Indeed, performing the calculation in the
Heisenberg representation we can write
\bea
\frac{\pa}{\pa x_j} \langle \cdots \rangle &=&
\langle \cdots \frac{\pa \Phi^-_{(a_j,b_j)}}{\pa x_j} \cdots \rangle =
\langle \cdots \frac{1}{\sqrt{2}} [L_{-1} - \bar{L}_{-1} ,
\Phi^-_{(a_j,b_j)}] \cdots \rangle \nonumber \\
&=&- \frac{i}{2\sqrt{2}} \langle \cdots  \left[ \{Q, G^{-}_{-\frac12} +i
\bar{G}^{-}_{-\frac12} \}, \Phi^-_{(a_j,b_j)} \right] \cdots \rangle =0
\ena
and
\bea
\frac{\pa}{\pa t_j} \langle \cdots \rangle &=&
\langle \cdots \frac{1}{\sqrt{2}} [L_{-1} + \bar{L}_{-1} ,
\Phi^-_{(a_j,b_j)}] \cdots \rangle \nonumber \\
&=& -\frac{i}{2\sqrt{2}} \langle \cdots  \left[ \{Q, G^{-}_{-\frac12} -i
\bar{G}^{-}_{-\frac12} \}, \Phi^-_{(a_j,b_j)} \right] \cdots \rangle =0
\ena
Thefore we can compute correlation functions for vertex operators
at the
same position and time or, in light--cone coordinates, with $z_1 =
z_2 =
\cdots = z_N$, $\bar{z}_1 = \bar{z}_2 = \cdots = \bar{z}_N$.

Additional restrictions come from the requirement of topological charge
conservation. Physically this
condition means that the correlation function is non--vanishing
only when solitons and antisolitons are present in equal number.
{}From eq. (4.10) we have
\EQ
\left[ {\cal T}^{+},\Phi^{-}_{(a_i,b_i)} \right] = \b^2 (a_i - b_i)
\Phi^-_{(a_i,b_i)}
\EN
Thus we need impose
\EQ
\sum_{i=1}^N {\cal T}_i^{+} \equiv \sum_{i=1}^N a_i -
\sum_{i=1}^N b_i = 0
\EN
This constraint and space--time independence allow to reduce all
topological correlators to one--point correlation functions
\EQ
{\cal F}_N = \langle 0|\Phi_a^{-} (z_1,\bar{z}_1) |0\rangle_{\rm top}
\EN
where we have set $a=\sum_{i=1}^N a_i$ and $(z_1,\bar{z}_1)$ is any point
in the two dimensional space--time.
Clearly it is the topological nature of the theory which prevents the
existence of non--trivial solitonic asymptotic states.

We compute now ${\cal F}_N$ using perturbation theory.
As previously discussed the $A^{(1)}(1,1)$ theory is obtained from
the $A(1,0)$ system by adding the perturbations in eqs. (\ref{310}),
(\ref{10}). Since the ${\cal S}_+$ term is $Q$--exact,
perturbation theory in
$g_+$ is trivial.
The relevant perturbation is ${\cal S}_-$ given in eq. (\ref{310}).
Thus we compute perturbatively in
$g_-$ and write
\EQ
{\cal F}_N = \langle 0 | \Phi^{-}_a(z_1,\bar{z}_1)~
e^{{\cal S}_-} |0\rangle_{\rm top}
\label{117}
\EN
where now $|0\rangle$ is the topological $A(1,0)$ vacuum
annihilated by the
$A(1,0)$ BRST charge explicitly given in eq. (5.2) with
$g_-=0$.

Having insured the conservation of the topological charge, we still need
impose that the $O(1,1)$
charge and the background charge
($\frac{1}{\b}$,$\frac{1}{\b}$) be balanced in the correlation function.
We examine first the $O(1,1)$ charge. As it usually
happens in standard $N=2$ topological theories \cite{b5,b16,b12},
the twisting procedure
generates a $O(1,1)$ anomaly, since we have now
\bea
J_0 = \left[ J_1, L_{-1}' \right]  \qquad ~ J_0^{\dag} = -\left[ J_{-1},
L_1' \right] = J_0 + \frac{c}{3} \nonumber \\
\bar{J}_0 = \left[ \bar{J}_1, \bar{L}_{-1}' \right]  \qquad ~
\bar{J}_0^{\dag} = -\left[ \bar{J}_{-1},
\bar{L}_1' \right] = \bar{J}_0 - \frac{c}{3}
\ena
where $c$ is the central charge of the $A(1,0)$ theory.
It follows that non zero contributions to the
topological correlation function in eq. (6.7)
at order M in the perturbative
expansion only arise if
\EQ
a -2M = \frac{c}{3}
\label{114}
\EN
where $(a, -a)$ and $(-2,2)$ are the $O(1,1)$ charges of the
vertex operator and the perturbation respectively.

It is convenient to compute the topological correlation functions
as $N=2$ correlators in the untwisted theory
and implement the $O(1,1)$ anomaly
with a vertex
carrying $(-\frac{c}{3},\frac{c}{3})$ charges at infinity.
The simplest operator which realizes this and is not BRST trivial, is the
local antichiral primary field
\EQ
\Phi^+_{\frac{c}{3}} \equiv e^{i\b \frac{c}{3} \phi^{+}}
\EN
Obviously its insertion in the correlation function breaks BRST invariance.
As discussed in Ref. \cite{b12} this can be cured defining
\bea
&&\langle 0 | \Phi^{-}_a(z_1,\bar{z}_1) | 0 {\rangle}_{\rm top} =
 \lim_{z_{2}, ~\bar{z}_{2} \to \infty}
\langle 0 |~ [(z_1- z_{2})(\bar{z}_1-\bar{z}_{2})]^{a}~
\Phi^{-}_a(z_1,\bar{z}_1)~
\Phi^+_{\frac{c}{3}}(z_{2},\bar{z}_{2}) ~e^{\hat{{\cal S}}_-} |
0{\rangle}
\nonumber \\
&&~~~~~~~~~~~
\label{11}
\ena
where
\EQ
\hat{{\cal S}}_- =
\frac{g_-}{2\pi \b^2}\int d^2z d^2\bar{\theta}~
[(z-z_{2})(\bar{z}-\bar{z}_{2})]^{-2}~ e^{-i\b \Psi^{-}}
\label{125}
\EN
Indeed it is easy to verify that the introduction of the $(z_2, \bar{z}_2)$
factors in the chiral vertex $\phi^{-}_a$ and in the ${\cal S}_-$
perturbation are such that the modified correlation function $\langle
\cdots \rangle$ is still $(z_1,\bar{z}_1)$--independent.
This can be shown
explicitly using a modified charge
\EQ
\hat{Q} = \int \frac{dz}{2\pi i} (z-z_2)
\left[ iG^{+}(z,\bar{z})+ \bar{G}^{+}(z,\bar{z}) \right]
\EN
which commutes with $\Phi^+_{\frac{c}{3}}$.
Taking the limit $z_2,~\bar{z}_2 \to \infty$ at the end of the calculation
one restores the standard BRST invariance of the topological correlation
function.

At this stage the calculation of $\cal{F}_N$ has been reduced to the
evaluation of a two--point function with the vertices
$\Phi^-_a$, $\Phi^+_{\frac{c}{3}}$ inserted at $(z_1,\bar{z}_1)$,
$(z_2,\bar{z}_2)$. In general a two--point correlation
function is invariant under Moebius transformations that
leave the points $z_1$ and $z_2$ fixed. Thus in order to obtain
a finite result we have to factor out
the infinite Moebius volume given by
\EQ
V_{\infty}\bar{V}_{\infty} = \int dz~
\frac{z_1-z_2}{|(z-z_1)(z-z_2)|} \int d\bar{z}~
\frac{ \bar{z}_1-\bar{z}_2}{|(\bar{z}-\bar{z}_1)(\bar{z}-\bar{z}_2)|}
\EN

We turn now to the problem of background charge balance which we need impose
in order to cancel the
dependence on the infrared cutoff $\m$ and obtain non--zero results
in the thermodynamic limit. To this end
screening operators must be inserted in eq. (6.11).
  For the $A(1,0)$ theory
there are one ``bosonic'' ${\cal U}$ and two ``fermionic'' ${\cal U}_{\pm}$
screening operators \cite{b16,b17}
which in terms of $N=2$ superfields are given by
\EQ
{\cal U} = \int d^2z d^4\theta ~e^{-\frac{i}{\b} (\Psi^{+} +\Psi^{-})}
\EN
\EQ
{\cal U}_+ = \int d^2z d^2\theta~ e^{i\b \Psi^{+}} \qquad \quad {\cal U}_- =
\int d^2z d^2\bar{\theta}~ e^{i\b \Psi^{-}}
\EN
The sum of the two fermionic screenings is the marginal
perturbation of the
$A(1,0)$ theory (see eq. (\ref{311})).
Since even the bosonic screening is a marginal operator one
could add ${\cal U}$ to the lagrangian without affecting its physical
content \cite{b18}. Thus the insertion of a
given
number of screening operators corresponds to the computation of the
correlation function at a given
order in the marginal perturbations.
Inserting
$(p-1)$ bosonic ${\cal U}$ screenings and $q_{\pm}$ fermionic ${\cal U}_{\pm}$
ones,
at order $M$ in $g_-$, the request of background charge conservation
gives
\bea
&& \b a  - \b M -(p-1) \frac{1}{\b} +
q_- \b =
-\frac{1}{\b} \nonumber \\
&& \b \frac{c}{3} - (p-1)\frac{1}{\b} + q_+ \b = -\frac{1}{\b}
\label{128}
\ena
The two equations are consistent with the $O(1,1)$ charge balance
(\ref{114})
only for
$q_+ - q_- = M$. Moreover setting $c = 3-\frac{6}{\b^2} $ in the second
equation we
find $ p= \b^2(q_- +M+1)$. Therefore non--zero correlation functions are
obtained only for Toda theories with $\b^2$ a positive integer.
Writing $\b^2 = k+2$, $k=1, 2, \cdots $ we obtain $c = \frac{3k}{k+2}$,
namely the central charge of the $A_{k+1}$ $N=2$ minimal models \cite{b19}.

We summarize here the results presented in this Section: the
calculation of the $N$--point topological correlation function,
at order $M$ in perturbation theory with respect to $g_-$, is
given by
\bea
{\cal F}_N^{(M)} &= & \langle 0 | \Phi^{-}_{(a_1,b_1)}(x_1,t_1)
\Phi^{-}_{(a_2,b_2)}(x_2,t_2) \cdots
\Phi^{-}_{(a_N,b_N)}(x_N,t_N) ~{\cal S}_-^M|0\rangle_{\rm top}
\nonumber\\
&&~~~~~~~~~~~~~~~~~~~~\nonumber\\
&=&\lim_{z_{2}, ~\bar{z}_{2} \to \infty} [ V_{\infty} \bar{V}_{\infty} ]^{-1}
{}~\langle 0 |~ [(z_1- z_{2})(\bar{z}_1-\bar{z}_{2})]^{a}~
\Phi^{-}_a(z_1,\bar{z}_1) ~
\Phi^{+}_{\frac{c}{3}}(z_{2},\bar{z}_{2}) \nonumber \\
&&~~~~~~~~~~~~~~~~~~~~~~~~~~~~~~~~~
{\hat{{\cal S}}_-^M}~{\cal U}^{p-1}~
{\cal U}_+^{q_+}~ {\cal U}_-^{q_-}~ | 0\rangle
\label{finn}
\ena
with ${\hat{\cal S}}_-$, ${\cal U}$, ${\cal U}_{\pm}$ and
$ V_{\infty} \bar{V}_{\infty}$ in eqs.(6.12, 6.15, 6.16, 6.14)
respectively and
\bea
&& c = 3-\frac{6}{\b^2}=\frac{3k}{k+2}Ê\qquad \qquad \qquad k=1, 2, \cdots
\nonumber\\
&&~~~~~~~~~\nonumber \\
&& \sum_{i=1}^N a_i ~~~= ~~~\sum_{i=1}^N b_i
{}~~~= ~~~a~~~=~~~ 2M+\frac{k}{k+2} \nonumber\\
&&~~~~~~~~~~~~~~\nonumber \\
&& q_+-q_-=M \qquad ~~~~~~~~~~\qquad p-1=(k+2)(q_+ +1) -1
\label{val}
\ena

\sect{The superspace calculation}

In this section we use the $N=2$ superspace formalism to compute
the simplest example of topological correlation functions in
eq. (\ref{finn}):
we choose the model with $c=1$ and perform the calculation
at zero order in the $g_-$ perturbation. Then from eq. (\ref{val}) we have
$k=1$, $\b^2=3$, $a=\frac {1}{3}$. Moreover the
minimum number of screening operators we
need insert in order to obtain a non--vanishing correlation
function
is two bosonic and zero fermionic ones. Thus we want to evaluate
\EQ
{\cal{F}}=\lim_{z_{2}, ~\bar{z}_{2} \to \infty}
[ V_{\infty} \bar{V}_{\infty} ]^{-1}
[(z_1- z_{2})(\bar{z}_1-\bar{z}_{2})]^{\frac{1}{3}}
 \langle~ :e^{\frac{i}{3} \b \phi^{-}(z_1,\bar{z}_1)}:~
:e^{\frac{i}{3} \b \phi^{+}(z_2,\bar{z}_2)}: {\cal U}~ {\cal U}~
\rangle
\label{corr}
\EN
For notational convenience we leave $\b$ unspecified, inserting
the value $\b^2 = 3$ in the final result. We perform first
the contraction of the two normal ordered exponentials
\EQ
 \langle~ :e^{\frac{i}{3} \b \phi^{-}(z_1,\bar{z}_1)}:~
:e^{\frac{i}{3} \b \phi^{+}(z_2,\bar{z}_2)}:~ {\cal U}~ {\cal U}~
\rangle
= e^{\frac{\b^2}{9}\log{|2(z_1-z_2)(\bar{z}_1 -\bar{z}_2)|}}
\langle ~: e^{\frac{i}{3} \b [\phi^{-}(z_1,\bar{z}_1)+
 \phi^{+}(z_2,\bar{z}_2)]}: ~{\cal U}~ {\cal U}~
\rangle
\label{115}
\EN
We note that being the theory interacting the screenings have a
nontrivial
dependence on the auxiliary fields. If we were to use a
component description and
eliminate the auxiliary fields via their field equations,
the calculation of the correlation function would become
hardly manageable.
A way to overcome this problem is to
perform
the computation off--shell directly in $N=2$ superspace.
We write then
\EQ
\langle ~:e^{\frac{i}{3}\b [\phi^{-}(z_1,\bar{z}_1)+
  \phi^{+}(z_2,\bar{z}_2)]}:~{\cal U}~ {\cal U}~  \rangle
= \left.
\langle ~:e^{\frac{i}{3}
\b  [\Psi^{-}(Z_1,\bar{Z}_1)+
 \Psi^{+}(Z_2,\bar{Z}_2)]}:~{\cal U}~ {\cal U}~  \rangle \right|
\label{13}
\EN
where we have introduced the notation $(Z,\bar{Z}) \equiv
(z,\theta,\bar{z},\bar{\theta})$
and $\left. \right|$ means evaluating the final result at $\theta_1 =
\bar{\theta}_1 = \theta_2 =\bar{\theta}_2 = 0$.
The $N=2$ superspace conventions we will adopt in the course
of the calculation are collected in Appendix A. There the reader can
also find the expression of the superfield propagators with
ultraviolet and infrared cutoffs explicitly shown.
Here we will not indicate the dependence on the cutoffs,
unless when necessary to prove cancellations of divergences.

We proceed organizing the calculation in eq. (\ref{13}) as follows: first we
compute the contraction of the two
screenings and then contract the result with the normal ordered
exponential.
Using the quantum--background method, we define quantum fields
$\xi$ as $\Psi
\rightarrow \Psi
+ \xi$ and compute
\bea
&& \underbrace{{\cal U}~ {\cal U}}
  = \int d^2z d^4\theta \int d^2z' d^4\theta'
 :\underbrace{  e^{-\frac{i}{\b}
(\xi^{+} + \xi^{-})(Z,\bar{Z})}:~~: e^{-\frac{i}{\b}
(\xi^{+} + \xi^{-})(Z',\bar{Z}')}}:  \nonumber \\
&& ~~~~~~~~~~~~~~~~~~~~~~~ : e^{-\frac{i}{\b}[
(\Psi^{+} + \Psi^{-})(Z,\bar{Z})+
(\Psi^{+} + \Psi^{-})(Z',\bar{Z}')]}:~~~~~~~~~~~~
\ena
where $\underbrace{~~~~}$ indicates the complete contraction of the
two operators. We obtain
\bea
&&:\underbrace{ e^{-\frac{i}{\b}
(\xi^{+} + \xi^{-})(Z,\bar{Z})}:~~: e^{-\frac{i}{\b}
(\xi^{+} + \xi^{-})(Z',\bar{Z}')}}:~~~~~~~~~~~~ \nonumber\\
&&~~~~~~~~~~=1+
\sum_{n=1}^{\infty} \left( -\frac{i}{\b} \right)^{2n} \frac{1}{(n!)^2}
\sum_{p=0}^n \left( \begin{array} {c} n \\ p \end{array} \right) \left(
\begin{array} {c} n \\ n-p \end{array} \right) p! (n-p)!Ê\nonumber\\
&&~~~~~~~~~~~~~~~~~~~~~
\langle~ \xi^+(Z,\bar{Z}) \xi^-(Z',\bar{Z'}) ~\rangle^p
\langle~ \xi^-(Z,\bar{Z}) \xi^+(Z',\bar{Z'}) ~\rangle^{n-p}
\label{121}
\ena
Using the explicit expression of the propagators as given
in eq. (\ref{A4}) one can perform the $D$--algebra in the
$~(n-1)$--loop order contribution which gives
\bea
&&\langle ~\xi^+(Z,\bar{Z})~ \xi^-(Z',\bar{Z'}) ~\rangle^p
\langle ~\xi^-(Z,\bar{Z}) ~\xi^+(Z',\bar{Z'}) ~\rangle^{n-p}~~~~~ \nonumber\\
&&~~~~~~~~=4\d^{(4)}(\theta-\theta') \left\{ -\frac{p(n-p)}{n-1}~~
(-\log|2(z-z')(\bar{z}-\bar{z}')|)^{n-1} ~
\pa \bar{\pa}(-\log|2(z-z')(\bar{z}-\bar{z}')|)
\right. \nonumber\\
&&~~~~~~~~~~~\left. +(-\log|2(z-z')(\bar{z}-\bar{z}')|)^n
\left[ D^2 \bar{D}^2 +
\frac{p}{n} (i\pa D_- \bar{D}_- -i\bar{\pa} D_+ \bar{D}_+)
-\frac{p(p-1)}{n(n-1)} \pa \bar{\pa} \right] \right\} \nonumber\\
&&~~~~~~~~~~~~~~
\label{prop}
\ena
where the spacetime and covariant spinor derivatives act on the
$(Z',\bar{Z'})$ variables. Inserting the above expression in eq. (\ref{121})
one can easily perform the sum over $p$ and use the identities in
eq. (A.16) to rewrite the result in the following form
\bea
&& :\underbrace{ e^{-\frac{i}{\b}
(\xi^{+} + \xi^{-})(Z,\bar{Z})}:~~: e^{-\frac{i}{\b}
(\xi^{+} + \xi^{-})(Z',\bar{Z}')}}:~~~~~  \nonumber\\
&&=1+\d^{(4)}(\theta-\theta')
\sum_{n=1}^{\infty} \left(\frac{2}{\b^2}\right)^n
\left \{ -\frac{1}{(n-1)!}
(\log|2(z-z')(\bar{z}-\bar{z}')|)^{n-1} ~\pa \bar{\pa}
(\log|2(z-z')(\bar{z}-\bar{z}')|) \right. \nonumber \\
&&~~~~~~~~~~~~\left. +\frac{1}{n!}
(\log|2(z-z')(\bar{z}-\bar{z}')|)^n \left[ D^2 \bar{D}^2 +
\bar{D}^2 D^2 +\bar{D}_+ D^2 \bar{D}_- +D_+ \bar{D}^2 D_-
 \right] \right \} \nonumber \\
&&~~~~~~~~~~~~~~~~~~~~~~~~~~~\nonumber \\
&&= 1+ \d^{(4)}(\theta-\theta')
\left\{ -\frac{2}{\b^2}|2(z-z')(\bar{z}-\bar{z}')|^{\frac{2}{\b^2}}
\pa \bar{\pa}
(\log|2(z-z')(\bar{z}-\bar{z}')|) \right. \nonumber \\
&&~~~~~~~~~~~~\left.
+ [|2(z-z')(\bar{z}-\bar{z}')|^{\frac{2}{\b^2}}-1]
\left[ D^2 \bar{D}^2 +
\bar{D}^2 D^2 +\bar{D}_+ D^2 \bar{D}_- +D_+ \bar{D}^2 D_-
 \right] \right\}
\label{fin}
\ena
We note that the term
\EQ
|2(z-z')(\bar{z}-\bar{z}')+\e^2|^{\frac{2}{\b^2}} ~\pa \bar{\pa}
(\log|2(z-z')(\bar{z}-\bar{z}')+\e^2|)
\label{sum}
\EN
has been obtained from the infinite sum of
contributions of the form
\EQ
(\log|2(z-z')(\bar{z}-\bar{z}')+\e^2|)^n ~\pa \bar{\pa}
\log|2(z-z')(\bar{z}-\bar{z}')+\e^2|
\label{term}
\EN
where we have explicitly indicated the dependence
on the ultraviolet cutoff $\e$.
Using the relation
\EQ
\pa \bar{\pa} \log|2(z-z')(\bar{z}-\bar{z}')|=2\pi i ~\d^{(2)}(z-z')
\EN
one can easily check that, while each term in eq. (\ref{term}) is
logarithmically
ultraviolet divergent, the final sum in eq. (\ref{sum}) vanishes
when the cutoff is removed. Going back to eq. (7.3)
we can then write
\bea
&&
\langle ~:e^{\frac{i}{3}
\b  [\Psi^{-}(Z_1,\bar{Z}_1)+
 \Psi^{+}(Z_2,\bar{Z}_2)]}:~{\cal U}~ {\cal U} ~ \rangle \nonumber \\
&&~~~= \int d^2z d^4 \theta~ f(Z,\bar{Z})
\int d^2z' d^4 \theta' ~f(Z',\bar{Z'}) +
\int d^2z d^2z' d^4 \theta~
\left[ |2(z-z')(\bar{z}-\bar{z}')|^{\frac{2}{\b^2}} -1\right] \nonumber\\
&&~~~~~~~~~~~~~~~~~f(Z,\bar{Z}) [D^2 \bar{D}^2 +
\bar{D^2} D^2 +\bar{D}_+ D^2 \bar{D}_- +D_+ \bar{D}^2 D_-]
 f(Z',\bar{Z'})
\ena
having defined
\EQ
f(Z, \bar{Z}) \equiv e^{\frac13 \langle
\Psi^{+}(Z,\bar{Z}) \Psi^{-}(Z_1,\bar{Z}_1) \rangle +
\frac13 \langle \Psi^{-}(Z,\bar{Z}) \Psi^{+} (Z_2,\bar{Z}_2)
\rangle}
\label{f}
\EN
Then we perform the $\theta$--integration. Inserting the
prefactor
in eq. (\ref{115}) we obtain
\bea
{\cal{I}}&\equiv&\langle~ :e^{i\b a \phi^{-}(z_1,\bar{z}_1)}: ~~
:e^{\frac{i}{3} \b  \phi^{+}(z_2,\bar{z}_2)}:~{\cal{U}}~
{\cal{U}} ~\rangle =
|2(z_1-z_2)(\bar{z}_1 -\bar{z}_2)|^{\frac{\b^2}{9}} ~~~~~~~~~~~  \nonumber \\
&&~~~~~ \left. \int d^2z d^2z'~
|2(z-z')(\bar{z}-\bar{z}')|^{\frac{2}{\b^2}}~
4D^2 \bar{D}^2 f(z,\theta,\bar{z},\bar{\theta}) \cdot
4\bar{D}^2 D^2 f(z',\theta,\bar{z}',\bar{\theta})
\right|
\label{16}
\ena
{}From the definition in eq. (\ref{f}), using the explicit expressions
of the propagators as given in eq. (\ref{A4}) of the Appendix,
we obtain
\bea
f(z,\theta,\bar{z},\bar{\theta})|_{\theta_1=\theta_2=0} &=&
e^{-\frac{1}{3} (1+i\theta_+\bar{\theta}_+ \pa)
(1-i\theta_-\bar{\theta}_- \bar{\pa}) \log|2(z-z_1)(\bar{z}-\bar{z}_1)|}
\nonumber \\
{}~&~&~~~~~~~e^{-\frac{1}{3}(1-i\theta_+\bar{\theta}_+ \pa)
(1+i\theta_-\bar{\theta}_- \bar{\pa}) \log|2(z-z_2)(\bar{z}-\bar{z}_2)|}
\label{ff}
\ena
It is now tedious but straightforward to evaluate the expressions
 $D^2 \bar{D}^2f$ and $\bar{D}^2 D^2 f$
at $\theta=\bar{\theta}=0$. Details  and intermediate steps of the explicit
calculation are given in Appendix B. Using the result in eq. (\ref{ii}), after
a
rescaling of all factors by $z_2\bar{z}_2$ we obtain
\bea
\lim_{z_{2}, ~\bar{z}_{2} \to \infty}
{\cal{I}} &\sim & \frac{2}{81}^{\frac{\b^2}{9}+\frac{2}{\b^2}-\frac{4}{3}}
(z_2\bar{z}_2)^{-\frac{1}{3}} \\
&& \int d^2zd^2z' \left| (z-z')(\bar{z}-\bar{z}')
\right|^{\frac{2}{\b^2}}
\frac{\left| z\bar{z} (z-1)(\bar{z}-1) z'\bar{z}' (z'-1)(\bar{z}'-1)
\right|^{-\frac13}}{z\bar{z} (z-1)(\bar{z}-1) z'\bar{z}'
(z'-1)(\bar{z}'-1)} \nonumber
\label{call}
\ena
We note that the integral rescales with the correct power in
$z_2\bar{z}_2$ in order to give a non--trivial contribution to the correlation
function as defined in eq. (\ref{corr}). Therefore for $\b^2 = 3$
we can finally write
\EQ
{\cal{F}}
= \frac{2^{-\frac{1}{3}}}{81} {\cal J}~\bar{{\cal J}}~
[{\cal V}_{\infty} \bar{{\cal V}}_{\infty}]^{-1}
\label{21}
\EN
where
\EQ
{\cal{V}}_{\infty} \equiv \lim_{z_{2}, ~\bar{z}_{2} \to \infty}
V_{\infty} =
\int \frac{dz}{|z(z-1)|}
\EN
and
\EQ
{\cal{J}} = \int dzdz'~ (z-z')^{\frac23} \frac{\left|
z(z-1)z'(z'-1) \right|^{-\frac13}}{z(z-1)z'(z'-1)}
\label{int}
\EN
with $\bar{{\cal{J}}}$ the same integral in the $\bar{z},~\bar{z}'$--variables.
We would like to remark that the correlation functions are
factorized as
products of holomorphic and antiholomorphic terms in spite of the
lack of explicit factorization in the chiral ring.

The last step is the evaluation of the integrals in eq. (\ref{int}).
The details are given in Appendix B. We report here the
final result: inserting the value of ${\cal J}$ from eq. (B.11)
into eq. (7.16) we obtain
\EQ
{\cal F} =  2^{-
\frac13}\left[ \frac{4}{3}
\frac{\G^2(\frac{2}{3})}{\G(\frac13)}\right]^2
\label{final}
\EN

\sect{Conclusions}

The affine Toda theories based on the Lie superalgebras $A^{(1)}(n,n)$
in two--dimensional Minkowski spacetime are nonunitary models
that can be twisted into a topological sector. They
can be obtained in a two--step procedure as perturbations of the
$N=2$ superconformal $A(n,n-1)$ Toda systems.

In this paper we have studied in detail the case $n=1$.
The action of the $A(1,0)$ model, $S_0$ given in eq. (3.2), contains
an interaction, $V_0$, which is a marginal perturbation of the
free theory, a necessary condition since the complete interacting theory
is still conformally invariant.
The affine theory, $A^{(1)}(1,1)$, is constructed by adding to $S_0$
the perturbation terms $S_-$ and $S_+$ in eqs. (3.11, 3.12).
We have shown that $S_-$ is a relevant perturbation while $S_+$
is $Q$--exact with respect to the BRST charge which defines
the topological version of the theory.

 Having identified the
chiral and antichiral rings, we concentrated on the computation
of the topological correlation functions. The physical, BRST--invariant
operators are primary chiral vertices with well--defined scale
dimensions, $O(1,1)$ charge and solitonic topological number.
When inserted into a correlation function they give a nonzero
result only if the various charges are conserved in the process.
This requirement with the additional request of background charge conservation
fixes the value of the central charge of the theory to be the one of
the $N=2$ $A_{k+1}$ minimal models. Moreover it
puts severe restrictions on the allowed quantum numbers of the vertices.
In particular solitons and antisolitons are always present in equal number.
In fact one finds that the calculation of an $N$--point correlation function
reduces to the evaluation of a one--point topological correlator,
independent of the solitonic number.

At this stage it is convenient to go back to the untwisted version
of the theory, implement the topological $O(1,1)$ anomaly with a vertex
carrying this anomalous charge at infinity and compute the correlation function
as in ordinary perturbation theory. Within this approach the general
form of the $N$--point correlator is given
in eq. (\ref{finn}). We emphasize that ${\cal F}_N^{(M)}$ is at order
M in the relevant $g_-$ perturbation and at order $(p-1)$, $q_{\pm}$ in
the marginal perturbations ${\cal U}$, ${\cal U}_{\pm}$ respectively.
The simplest correlator for the $c=1$ model and $M=0$ has been
computed
using a $N=2$ superspace approach. Superfield techniques simplify the
algebra significantly so that the calculation of higher--order
corrections in the $g_-$ perturbation might be attempted.

We have considered a theory in Minkowski spacetime. Obviously the
same procedure would work equally well in standard $N=2$
supersymmetric theories formulated in euclidean space. Since our
calculation consists essentially in the evaluation of a two--point
correlator, it would be interesting to apply the $N=2$ superspace
method to the calculation of the Zamolodchikov chiral--antichiral
metric \cite{b21}.

\vskip 15pt
\noindent
We acknowledge useful conversations with E. Montaldi, M. Pernici, P. Soriani
and N. Warner.

\appendix
\sect{}

In this appendix we list our notations and conventions.
We work in $N=2$ superspace described by light--cone coordinates
\EQ
z = \frac{x^0+x^1}{\sqrt{2}} \qquad \quad \bar{z} = \frac{x^0-
x^1}{\sqrt{2}}
\EN
where $x^0 \equiv t$ and $x^1$ are coordinates in Minkowski space
with signature $g_{\mu \nu} = diag(1,-1)$,
and spinor coordinates $\theta_{\pm}$, $\bar{\theta}_{\pm}$
satisfying
the following conjugation rules
\EQ
\theta_+^{\ast} = -\bar{\theta}_+ \qquad \quad
\theta_-^{\ast} = \bar{\theta}_-
\EN
We define $N=2$ supercovariant derivatives
\bea
D_+ &=& \frac{1}{\sqrt{2}} \left( \frac{\partial}{\partial \theta_+} +
i
\bar{\theta}_+
\partial \right) \qquad \bar{D}_+ = \frac{1}{\sqrt{2}} \left(
\frac{\partial}{\partial \bar{\theta}_+} + i \theta_+ \partial \right)
\nonumber \\
D_- &=& \frac{1}{\sqrt{2}} \left( \frac{\partial}{\partial \theta_-} -
i
\bar{\theta}_- \bar{\partial} \right)
\qquad \bar{D}_- = \frac{1}{\sqrt{2}} \left( \frac{\partial}{\partial
\bar{\theta}_-}
- i\theta_- \bar{\partial} \right)
\ena
where
\EQ
\partial \equiv \pa_z = \frac{1}{\sqrt{2}}
\left(\partial_0 + \partial_1 \right) \qquad
\bar{\partial} \equiv \pa_{\bar{z}} =
\frac{1}{\sqrt{2}} \left(\partial_0 - \partial_1 \right)
\EN
They satisfy the algebra
\EQ
\{ D_+,\bar{D}_+ \} = i \partial \qquad \quad \{D_-,\bar{D}_- \} = -i
\bar{\partial}
\label{118}
\EN
Often we make use of the following notation
\EQ
\theta_+ \theta_-\equiv \theta^2 \qquad \qquad
\bar{\theta}_+ \bar{\theta}_- \equiv \bar{\theta}^2 \qquad \qquad \qquad
D_+D_-\equiv D^2 \qquad \qquad \bar{D}_+ \bar{D}_-
\equiv\bar{D}^2
\EN
$N=2$ chiral and antichiral superfields $\Psi$ and
$\bar{\Psi} \equiv \Psi^{\ast}$ are subject
to the constraints
\EQ
\bar{D}_{\pm} \Psi = 0 \qquad \quad D_{\pm} \bar{\Psi} = 0
\EN
and they can be written as
\bea
\Psi(\theta_+,\theta_-,\bar{\theta}_+,\bar{\theta}_-) &=& e^{
i\theta_+
\bar{\theta}_+\partial  - i\theta_-\bar{\theta}_-\bar{\partial} }
\Psi(\theta_+,
\theta_-) \nonumber \\
\bar{\Psi}(\theta_+,\theta_-,\bar{\theta}_+,\bar{\theta}_-) &=& e^{
-i\theta_+
\bar{\theta}_+\partial  + i\theta_-\bar{\theta}_-\bar{\partial} }
\bar{\Psi}
(\bar{\theta}_+,\bar{\theta}_-)
\ena
where
\bea
\Psi(\theta_+,\theta_-) &=& \phi + \frac{1}{\sqrt{2}}\theta_+
\bar{\psi} +
\frac{1}{\sqrt{2}}\theta_- \psi +
\theta_+\theta_- F \nonumber \\
\bar{\Psi}(\bar{\theta}_+,\bar{\theta}_-) &=& \bar{\phi} +
\frac{1}{\sqrt{2}}
\bar{\theta}_+ \bar{\psi}^{\ast} - \frac{1}{\sqrt{2}}\bar{\theta}_-
\psi^{\ast}
+\bar{\theta}_+\bar{\theta}_- \bar{F}
\label{A1}
\ena
The convention for complex cojugation on the product of two
fermions
is $(\psi_1 \psi_2 )^{\ast} = \psi_2^{\ast} \psi_1^{\ast}$.
We have defined $\phi$ ($\bar{\phi} = \phi^{\ast}$) to be a
complex scalar field,
$F$ ($\bar{F} = F^{\ast}$) the auxiliary field and $\psi$,
$\bar{\psi}$
the components of a Dirac spinor
\EQ
\psi =  \frac{1}{2^{1/4}} \left(\begin{array} {c} \psi \\
\bar{\psi}
\end{array} \right)
\qquad \bar{\psi} \equiv \psi^{\ast} \gamma_0 = \frac{1}{2^{1/4}}
\left( \bar{\psi}^{\ast}~~ \psi^{\ast} \right)
\EN
with respect to the two dimensional Dirac basis
\EQ
\gamma_0 = \left( \begin{array} {cc}
0 & 1 \\
1 & 0
\end{array} \right)
\qquad \gamma_1 = \left( \begin{array} {cc}
{}~0 & 1 \\
-1 & 0
\end{array} \right) \qquad C = \gamma_1
\EN
where $C$ is the charge conjugation matrix.
Majorana--Weyl spinors satisfy the reality conditions $\psi^{\ast} =
\psi$,
$\bar{\psi}^{\ast} = -\bar{\psi}$.

We consider a general $N=2$ superspace action for $n$ superfields
\EQ
\frac{1}{2\pi} \int d^2z d^4 \theta ~K_{ij} \Psi_i \bar{\Psi}_j
+ \frac{1}{2\pi} \int d^2z d^2\theta ~W(\Psi) + \frac{1}{2\pi} \int
d^2z
d^2 \bar{\theta}~ W(\bar{\Psi})
\label{A2}
\EN
where $K_{ij}$ is an invertible, symmetric, real $n \times n$
constant matrix
and $W$ is the superpotential. The integration measure is defined
as
$d^2z = dz d\bar{z}$, $d^2\theta = d\theta_+ d\theta_- \rightarrow
2D^2$,
$d^2\bar{\theta} =
d\bar{\theta}_+ d\bar{\theta}_- \rightarrow 2\bar{D}^2$ and
$d^4\theta = d^2\theta
d^2\bar{\theta} \rightarrow 4D^2\bar{D}^2$.
{}From the action in eq. (\ref{A2}) we obtain
the chiral--antichiral superfield propagator (with
$Z \equiv (z,\theta_+,\theta_-)$, $\bar{Z} \equiv (\bar{z},
\bar{\theta}_+,
\bar{\theta}_-$))
\EQ
\langle \Psi_i(Z,\bar{Z}) \bar{\Psi}_j(Z',\bar{Z}') \rangle =
-4~K_{ij}^{-1}\bar{D}^2 D^2 \delta^{(4)}(\theta -
\theta')
\log{|\m^2[2(z-z')(\bar{z}-\bar{z}')+\e^2]|}
\label{A4}
\EN
where
\EQ
\delta^{(4)}(\theta -\theta') = (\bar{\theta} -\bar{\theta}')^2
(\theta - \theta')^2
\EN
and $\m$, $\e$ are infrared and ultraviolet cutoffs respectively.
Standard superspace techniques greatly simplify the computation
of Green's functions.
Typically the final result for a perturbative loop contribution
is always local in the $\theta$--variables once
the $D$--algebra has been completed. In so doing one makes use
of the commutation algebra in eq. (\ref{118}) and of the relation
\EQ
4\delta^{(4)}(\theta -\theta') D^2\bar{D}^2 \delta^{(4)}(\theta -\theta')
=\delta^{(4)}(\theta -\theta')
\EN
We list here some identities that we have employed repeatedly in the
superspace calculation presented in Section 7:
\bea
D^2 \bar{D}^2 &=&\bar{D}^2 D^2 -i\bar{\pa}\bar{D}_+ D_+
+i \pa \bar{D}_- D_- - \pa \bar{\pa} \nonumber\\
&=&\bar{D}^2 D^2 +i\bar{\pa}D_+ \bar{D}_+
-i \pa D_- \bar{D}_- +\pa \bar{\pa}Ê\nonumber \\
&&~~~~~~~~~~~~~~~~~~~\nonumber \\
D^2 \bar{D}^2 +\bar{D}^2 D^2&=&\bar{D}_+ D^2 \bar{D}_-
+D_+ \bar{D}^2 D_- -\pa \bar{\pa}
\label {ident}
\ena
{}~~~~~~~~~~~~~~~~\\

Finally for a spin--1 conserved current $J_{\mu}$ ($ \partial^{\mu}
J_{\mu} =
0 $) we define light--cone components
\EQ
J = \frac{J_0 + J_1}{\sqrt{2}} \qquad \quad \tilde{J} =
\frac{J_0-J_1}{\sqrt{2}}
\label{A3}
\EN
satisfying the conservation equations $\bar{\partial} J = -\partial
\tilde{J}$. The corresponding conserved charge is
$ \int \frac{dx}{2\pi i \sqrt{2}} ( J +\tilde{J} )$ and in general we
define modes as
\EQ
J_n = \int \frac{dx}{2\pi i\sqrt{2}} \left(
\frac{x}{\sqrt{2}} \right)^n (J +\tilde{J})~~~~~~~~~~~~n=0, \pm 1,
\pm 2,\cdots
\EN
These definitions are easily extended to higher spin--conserved
currents.
Therefore, for the supersymmetry currents
$G_{\mu}$ and $\bar{G}_{\mu}$ we define the light--cone
components as
\bea
G &=& \frac{G_0 + G_1}{\sqrt{2}}
\qquad \quad \tilde{G} =
\frac{G_0-G_1}{\sqrt{2}} \nonumber \\
\bar{G} &=& \frac{\bar{G}_0 - \bar{G}_1}{\sqrt{2}}
\qquad \quad \tilde{\bar{G}} =
\frac{\bar{G}_0+\bar{G}_1}{\sqrt{2}}
\ena
and the corresponding modes
\bea
G_{n+\frac12} &=& \int \frac{dx}{2\pi i \sqrt{2}}
\left(\frac{x}{\sqrt{2}}\right)^{n+1} (G +
\tilde{G}) \nonumber \\
\bar{G}_{n+\frac12} &=& \int \frac{dx}{2\pi i \sqrt{2}}
\left(\frac{x}{\sqrt{2}}\right)^{n+1}(\bar{G} +
\tilde{\bar{G}})
\ena
The supersymmetry charges are obtained setting $n=-1$.
For the spin--2 stress energy tensor we define
\bea
T &\equiv& T_{++} = T_{00} + T_{11} + 2T_{01} \nonumber \\
\bar{T} &\equiv& T_{--} = T_{00} + T_{11} - 2T_{01} \nonumber \\
\tilde{T} &\equiv& T_{+-} = T_{00} - T_{11}
\ena
The modes corresponding to the $T$ component are
\EQ
L_n = \int \frac{dx}{2\pi i \sqrt{2}}
\left(\frac{x}{\sqrt{2}}\right)^{n+1} T
\EN
Analogous definitions hold for $\bar{T}$ and $\tilde{T}$.

\sect{}

In this appendix we present some details of the calculation
of the ${\cal{I}}$ and ${\cal{J}}$ integrals in Section 7.
Using the expression in eq. (\ref{ff}) we can compute
the quantity in eq. (\ref{16})
\bea
\cal{I} &\equiv& |2(z_1-z_2)(\bar{z}_1 -\bar{z}_2)|^{\frac{\b^2}{9}}
\left. \int d^2z d^2z'
{}~|2(z-z')(\bar{z}-\bar{z}')|^{\frac{2}{\b^2}} \right.\nonumber \\
&~&~~~~~~~~~~~~~~~~~~~~~~~~~ \left.
4D^2 \bar{D}^2 f(z,\theta,\bar{z},\bar{\theta}) \cdot
4\bar{D}^2 D^2 f(z',\theta,\bar{z}',\bar{\theta})
\right| \nonumber \\
&&~~~~~~~\nonumber \\
&=& 2^{\frac{\b^2}{9}+\frac{2}{\b^2}-\frac{4}{3}}
[(z_1-z_2)(\bar{z}_1-\bar{z}_2)]^{\frac{\b^2}{9}} \int d^2zd^2z'
\nonumber \\
&& \left| (z-z')(\bar{z}-\bar{z}') \right|^{\frac{2}{\b^2}} \left|
(z-z_1)(\bar{z}-\bar{z}_1)(z'-z_1)(\bar{z}'-\bar{z}_1) (z-
z_2)(\bar{z}-
\bar{z}_2) (z'-z_2)(\bar{z}'-\bar{z}_2)\right|^{-\frac{1}{3}}
\nonumber \\
&&~~~~~~~~~~~~~~~\nonumber \\
&& \left[ \frac13 \pa_{\bar{z}} \frac{1}{z-z_1} + \frac13 \pa_{\bar{z}}
\frac{1}{z-z_2} + \frac{1}{9}  \frac{1}{\bar{z}-\bar{z}_1}
\frac{1}{z-z_2} + \frac{1}{9}\frac{1}{z-z_1} \frac{1}{\bar{z}-\bar{z}_2}
\right. \nonumber \\
&&~~~~~~~~~~~~~~~~~~~~~~~~~~~~~~~~~~~~~~~~~~~~~~~~~~~
\left. - \frac{1}{9} \frac{1}{z-z_1}\frac{1}{\bar{z}-\bar{z}_1} -
\frac{1}{9} \frac{1}{z-z_2}\frac{1}{\bar{z}-\bar{z}_2} \right]
\nonumber \\
&& \left[ \frac13 \pa_{\bar{z}'} \frac{1}{z'-z_1} + \frac13 \pa_{\bar{z}'}
\frac{1}{z'-z_2} + \frac{1}{9}  \frac{1}{\bar{z}'-\bar{z}_1}
\frac{1}{z'-z_2} + \frac{1}{9}\frac{1}{z'-z_1} \frac{1}{\bar{z}'-\bar{z}_2}
\right. \nonumber \\
&&~~~~~~~~~~~~~~~~~~~~~~~~~~~~~~~~~~~~~~~~~~~~~~~~~~~
 \left. - \frac{1}{9} \frac{1}{z'-z_1}\frac{1}{\bar{z}'-\bar{z}_1} -
\frac{1}{9} \frac{1}{z'-z_2}\frac{1}{\bar{z}'-\bar{z}_2} \right]
\ena
We note that each term containing
 $\pa_{\bar{z}}$ and/or $\pa_{\bar{z}'}$ would lead to a divergent
contribution and then should be regularized appropriately.
Actually, by integration by parts, it is quite easy to show that
these terms cancel out completely.
Thus we end up with the following expression
\bea
&& {\cal{I}} = \frac{2}{81}^{\frac{\b^2}{9}+\frac{2}{\b^2}-\frac{4}{3}}
[(z_1-z_2)(\bar{z}_1-\bar{z}_2)]^{\frac{\b^2}{9}}
\int d^2z d^2z' \nonumber \\
&& \left| (z-z')(\bar{z}-\bar{z}') \right|^{\frac{2}{\b^2}} \left|
(z-z_1)(\bar{z}-\bar{z}_1)(z'-z_1)(\bar{z}'-\bar{z}_1) (z-
z_2)(\bar{z}-
\bar{z}_2) (z'-z_2)(\bar{z}'-\bar{z}_2)\right|^{-\frac{1}{3}}
\nonumber \\
&& \left( \frac{1}{\bar{z}-\bar{z}_1}\frac{1}{z-z_2} +
\frac{1}{z-z_1} \frac{1}{\bar{z}-\bar{z}_2} -
\frac{1}{z-z_1}\frac{1}{\bar{z}-\bar{z}_1} -
\frac{1}{z-z_2}\frac{1}{\bar{z}-\bar{z}_2} \right) \nonumber \\
&& \left( \frac{1}{\bar{z}'-\bar{z}_1}
\frac{1}{z'-z_2} + \frac{1}{z'-z_1} \frac{1}{\bar{z}'-\bar{z}_2}  -
\frac{1}{z'-z_1}\frac{1}{\bar{z}'-\bar{z}_1} -
\frac{1}{z'-z_2}\frac{1}{\bar{z}'-\bar{z}_2} \right)
\ena
which can be rewritten as
\bea
&& {\cal{I}}=\frac{2}{81}^{\frac{\b^2}{9}+\frac{2}{\b^2}-\frac{4}{3}}
[(z_1-z_2)(\bar{z}_1-\bar{z}_2)]^{\frac{\b^2}{9}} \int d^2zd^2z'
\nonumber\\
&& \left| (z-z')(\bar{z}-\bar{z}') \right|^{\frac{2}{\b^2}} \left|
(z-z_1)(\bar{z}-\bar{z}_1)(z'-z_1)(\bar{z}'-\bar{z}_1) (z-
z_2)(\bar{z}-
\bar{z}_2) (z'-z_2)(\bar{z}'-\bar{z}_2)\right|^{-\frac{1}{3}}
\nonumber \\
&&~~~~~~~~\nonumber \\
&& \left( \frac{1}{z-z_1} -\frac{1}{z-z_2}\right) \left(
\frac{1}{\bar{z}-\bar{z}_1} -\frac{1}{\bar{z}-\bar{z}_2}\right)
\left( \frac{1}{z'-z_1} -\frac{1}{z'-z_2}\right) \left(
\frac{1}{\bar{z}'-\bar{z}_1} -\frac{1}{\bar{z}'-\bar{z}_2}\right)
\label{ii}
\ena
Changing variables $z\rightarrow z/z_2$, $z'\rightarrow z'/z_2$,
it is now straightforward to obtain the asymptotic expression
in eq. (7.15).\\
{}~~~~~~~~~~~~~~~~~~~~~~~~~~\\

Finally we compute the integral in eq. (\ref{int}).
We start by considering the general expression
\EQ
{\cal{J}} = \int_{-\infty}^{+\infty} dz dz' |z-z'|^{-2b}
\frac{|z(z-1)z'(z'-1)|^b}
{z(z-1)z'(z'-1)}
\EN
$b$ being of the form $b=\frac{p}{2q+1}$, $p,q$ integers such that
the
integral is convergent. Then we define the integral for any value of
$p,q$ by
analytic continuation.

We compute ${\cal{J}}$ trying to factorize the infinite volume
${\cal{V}}_{\infty}= \int
\frac{dz}
{|z(z-1)|}$ of the one parameter Moebius transformations. Therefore
we write
\EQ
{\cal{J}} = {\int}_{-\infty}^{+\infty} \frac{dz}{|z(z-1)|} [z(z-1)]^b ~
{\int}_{-\infty}^{+\infty} dz' (z-z')^{-2b} \frac{|z'(z'-1)|^b}{z'(z'-1)}
\label{B1}
\EN
and evaluate the $z'$--integral first. We split the integration in
three intervals
\bea
{\cal{J}}' &\equiv& {\int}_{-\infty}^{+\infty} dz' (z-z')^{-2b}
\frac{|z'(z'-1)|^b}{z'(z'-1)} \nonumber \\
&=& {\int}_{-\infty}^0 dz' (z-z')^{-2b} [z'(z'-1)]^{b-1} - {\int}_0^1 dz'
(z-z')^{-2b} [z(1-z)]^{b-1} \nonumber \\
&~&+ {\int}_1^{+\infty} dz' (z-z')^{-2b} [z'(z'-1)]^{b-1}
\ena
Performing the change of variables $z' = \frac{u}{u-1}$ in the first
integral
and $z'=\frac{1}{u}$ in the third one we write the three
integrals in the form
\EQ
\int_0^1 dx~ x^{\l -1}(1-x)^{\mu -1} (1-{\cal H}(z) x)^{-\nu}
\EN
where ${\cal H}$ indicates any function of $z$.
Using eq. (3.197.3) page 286
in Ref. \cite{b20} we obtain the following expression
\EQ
{\cal J}' = B(b,1) F(2b,1,1+b;1-z) - z^{-2b} B(b,b) F(2b,b,2b;\frac{1}{z}) +
B(1,b) F(2b,1,1+b;z)
\EN
where $B(a,b) = \frac{\G(a)\G(b)}{\G(a+b)}$ and
$F(a,b,c;z)$ is the hypergeometric function.
Using the relations for analytic continuation of the function $F$
we can write
\bea
F(2b,1,1+b;1-z) &=& -F(2b,1,1+b;z)
+ z^{-b} \frac{\G(1+b)\G(b)}{\G(2b)} F(1-b,b,1-b;z)
\nonumber \\
&=& -F(2b,1,1+b;z) + \frac{\G(1+b)\G(b)}{\G(2b)} [z(z-1)]^{-b}
\nonumber \\
F(2b,b,2b;\frac{1}{z}) &=& z^b(z-1)^{-b}
\ena
Summing all the contributions, for the $z'$ integration we obtain
\EQ
{\cal J}' = [z(z-1)]^{-b} \cdot (-2) \frac{\G(b)^2}{\G(2b)}
\EN
Inserting the result in eq. (\ref{B1}) and simplifying the factor
$[z(z-1)]^b$ we can finally write
\EQ
{\cal J} = -2 \frac{\G(b)^2}{\G(2b)} \cdot {\cal V}_{\infty}
\EN


\newpage
\begin{thebibliography}{99}
\bibitem{b1} A.E. Arinshtein, V.A. Fateev and A.B.
Zamolodchikov,
\PL{B87} (1979) 389; \\
J.L. Cardy and G. Mussardo, \PL{B225} (1989) 275; \\
P.G.O. Freund, T.R. Klassen and E. Melzer, \PL{B229} (1989) 243; \\
P. Christe and G. Mussardo, \NP{B330} (1990) 465; \\
C. Destri and H.J. de Vega, \PL{B233} (1989) 336; \\
G. Mussardo and G. Sotkov, in ``Recent developments in Conformal Field
Theories'', Trieste, Italy, October 2--4, 1989; \\
T.R. Klassen and E. Meltzer, \NP{B338} (1990) 485; \\
H.W. Braden, E. Corrigan, P.E. Dorey and R.
Sasaki, \PL{B227} (1989) 441; \NP{B338} (1990) 689; \\
G.W. Delius, M.T. Grisaru, S. Penati and D. Zanon, \PL{B256} (1991) 164;
\NP{B359} (1991) 125; \\
C. Destri, H.J. De Vega and V.A. Fateev, \PL{B256} (1991) 173; \\
G.W. Delius, M.T. Grisaru and D. Zanon, \PL{B277} (1992) 414;
\NP{B382} (1992) 365.
\bibitem{b2} A.B. Zamolodchikov and Al.B. Zamolodchikov, Ann.
Phys. {\bf 120} (1979) 253.
\bibitem{b3} F.A. Smirnov, J. Phys. {\bf A17} (1984) L873; {\bf A19}
(1986) 575; \\
A.N. Kirillov and F.A. Smirnov, \PL{B198} (1987) 506; Int. J. Mod. Phys.
{\bf A3} (1988) 731; \\
A. Fring, G. Mussardo and P. Simonetti, \NP{B393} (1993) 413;
\PL{B307} (1993) 83; \\
A. LeClair, ``Spectrum generating Affine Lie algebras and correlation functions
in Massive Field Theory'', preprint CLNS--93--1220 (May 1993), hepth/9305110;
\\
A. LeClair and C. Efthimiou, ``Particle--field duality and form--factors
from vertex operators'', preprint CLNS--93--1263 (Dec 1993), hepth/9312121.
\bibitem{b4} E. Witten, \CMP{117} (1988) 353; \CMP{118} (1988) 411;
\NP{B340} (1990) 281.
\bibitem{b5} C. Vafa, Mod. Phys. Lett. {\bf A6} (1991) 337; \\
R. Dijkgraaf, E. Verlinde and H. Verlinde, Nucl. Phys.
{\bf B352} (1991) 59.
\bibitem{b6} B. Dubrovin, ``Integrable systems and classification of
2--dimensional topological field theories'', S.I.S.S.A. preprint 162/92/FM,
September 1992, hepth/9209040; \\
S. Cecotti, L. Girardello and A. Pasquinucci, \NP{B328} (1989) 701.
\bibitem{Ol} M.A. Olshanetsky, \CMP{88} (1983) 63.
\bibitem{b7} A. Gualzetti, S. Penati and D. Zanon, \NP{B398} (1993) 622.
\bibitem{b8} S. Penati and D. Zanon, in ``String theory, quantum
gravity and the unification of the fundamental interactions'',
eds. M. Bianchi et al., World Scientific (1993), 450.
\bibitem{b9} S. Penati, M. Pernici and D. Zanon, \PL{B309} (1993) 304.
\bibitem{b10} J. Evans and T. Hollowood, Phys. Lett. {\bf B293} (1992) 100; \\
J. Evans, \NP{B390} (1993) 225.
\bibitem{b11} H.C. Liao, D. Olive and N. Turok, \PL{B298} (1993) 95.
\bibitem{b12} N. Warner, ``N=2 supersymmetric integrable models and
topological field theories'', Proceedings of the Summer School on High Energy
Physics and Cosmology, Trieste, Italy, June 15th--July 3rd, 1992,
hep--th/9301088.
\bibitem{b13} D. Bernard and A. LeClair, \CMP{142} (1991) 99.
\bibitem{b14} S.J. Chang and R. Rajaraman, \PL{B313} (1993) 59; ``Chiral
vertex operators in off--conformal theory: the sine--Gordon example'',
hepth/9311107.
\bibitem{b15} M. Bershadsky, W. Lerche, D. Nemeschansky and N. Warner,
\NP{B401} (1993) 304.
\bibitem{b16} K. Li, \NP{B354} (1991) 711.
\bibitem{b17} M. Yu and H.B. Zheng, \NP{B288} (1987) 275; \\
M. Kato and S. Matsuda, \PL{B184} (1987) 184; \\
G. Mussardo, G. Sotkov and M. Stanishkov, Int. J. Mod. Phys. {\bf A4}
(1989) 1135; \\
K. Ito, \PL{B230} (1989) 71; \NP{B332} (1990) 566.
\bibitem{b18} H.C. Liao and P. Mansfield, \PL{B255} (1991) 237.
\bibitem{b19} A.B. Zamolodchikov and V.A. Fateev, Sov. Phys. JETP {\bf
63} (1985) 913; \\
P. di Vecchia, J.L. Petersen and H.B. Zheng, \PL{B162} (1986) 327; \\
P. di Vecchia, J.L. Petersen and M. Yu, \PL{B172} (1986) 211; \\
W. Boucher, D. Friedan and A. Kent, \PL{B172} (1986) 316; \\
P. di Vecchia, J.L. Petersen, M. Yu and H.B. Zheng, \PL{B174} (1986) 280.
\bibitem{b20} I.S. Gradshteyn and I.M. Ryzhik, ``Table of Integrals,
Series and Products'', Academic Press (1965).
\bibitem{b21} S. Cecotti and C. Vafa, \NP{B367} (1991) 359.

\end{thebibliography}
\end{document}